\documentclass[fleqn,usenatbib]{mnras}
    
\usepackage{newtxtext,newtxmath}
\usepackage[T1]{fontenc}
\usepackage{ae,aecompl}
\usepackage{multirow}
    
\usepackage{graphicx}	 
\usepackage{amsmath}	
\usepackage{amssymb}	
\newcommand{\hii}{H$_2\,$}	
\newcommand{\gammah}{$\gamma_{\mathrm{H_2}}$}   
 
\usepackage{color}
\usepackage[normalem]{ulem}

\definecolor{darkgreen}{rgb}{0.13, 0.55, 0.13}

\newcommand{\aref}[1]{\hyperref[#1]{Appendix~\ref{#1}}}

\title[Magnetic Fields and Population III IMF]{The importance of magnetic fields for the initial mass function of the first stars}

\author[Sharda, Federrath and Krumholz]{
Piyush Sharda$^{1,2}$\thanks{piyush.sharda@anu.edu.au (PS)},
Christoph Federrath$^{1,2}$\thanks{christoph.federrath@anu.edu.au (CF)},
and Mark R. Krumholz$^{1,2,3,4}$\thanks{mark.krumholz@anu.edu.au (MRK)}
\\
$^{1}$Research School of Astronomy and Astrophysics, Australian National University, Canberra, ACT 2611, Australia\\
$^{2}$Australian Research Council Centre of Excellence for All Sky Astrophysics in 3 Dimensions (ASTRO 3D), Australia\\
$^{3}$Universit\"at Heidelberg, Zentrum f\"ur Astronomie, Institut f\"ur Theoretische Astrophysik, 69120 Heidelberg, Germany\\
$^{4}$Max Planck Institute for Astronomy, K\"onigstuhl 17, 69117 Heidelberg, Germany
}

\date{Accepted 2020 June 30. Received 2020 June 30; in original form 2020 February 26}
    
\pubyear{2020}

\begin{document}
\label{firstpage}
\pagerange{\pageref{firstpage}--\pageref{lastpage}}
\maketitle

\begin{abstract}
Magnetic fields play an important role for the formation of stars in both local and high-redshift galaxies. Recent studies of dynamo amplification in the first dark matter haloes suggest that significant magnetic fields were likely present during the formation of the first stars in the Universe at redshifts of 15 and above. In this work, we study how these magnetic fields potentially impact the initial mass function (IMF) of the first stars. We perform 200 high-resolution, three-dimensional (3D), magneto-hydrodynamic (MHD) simulations of the collapse of primordial clouds with different initial turbulent magnetic field strengths as predicted from turbulent dynamo theory in the early Universe, forming more than 1100 first stars in total. We detect a strong statistical signature of suppressed fragmentation in the presence of strong magnetic fields, leading to a dramatic reduction in the number of first stars with masses low enough that they might be expected to survive to the present day. Additionally, strong fields shift the transition point where stars go from being mostly single to mostly multiple to higher masses. However, irrespective of the field strength, individual simulations are highly chaotic, show different levels of fragmentation and clustering, and the outcome depends on the exact realisation of the turbulence in the primordial clouds. While these are still idealised simulations that do not start from cosmological initial conditions, our work shows that magnetic fields play a key role for the primordial IMF, potentially even more so than for the present-day IMF.
\end{abstract}





\section{Introduction}
\label{s:intro}
Magnetic fields are ubiquitous in the Universe, and have a major impact on the behaviour of objects whose sizes range from planetary cores \citep{2003E&PSL.208....1S,2010SSRv..152...23B} to the intracluster medium \citep{2002ARA&A..40..319C,2013A&ARv..21...62D}. The relevance of magnetic fields in contemporary star formation has been extensively studied in theory, simulations and observations (see reviews by \citealt{1987ARA&A..25...23S,2007ARA&A..45..565M,2012ARA&A..50...29C,2017ARA&A..55..111H,2019FrASS...6....7K,2019FrASS...6....5H}). Nonetheless, several questions that concern magnetic fields still remain unanswered; for example, how are they generated? How are they amplified or dissipated? How are they sustained and how do they evolve across different scales? And importantly, how do they affect the stellar initial mass function (IMF)?

Magnetic fields have also been proposed to be of importance in the high-redshift Universe \citep{2006AN....327..505Z,2008ApJ...676...70K,2008Natur.454..302B}, especially during the formation of the first generation of stars around $z \sim 20-30$ \citep{2013RPPh...76k2901B,2019ffbh.book...67K}. However, the magnetic field strength at such high redshifts is extremely difficult to measure and to constrain; moreover, any constraints that arise can only inform us of the field strength and/or topology on scales much larger than that of molecular clouds where actual star formation takes place. Several physical processes can lead to the production of a magnetic field in the early Universe \citep{2001PhR...348..163G,2016RPPh...79g6901S}, including cosmological phase transitions ($10^{-40}\,\mathrm{s}$ after the Big Bang) that drive electric currents \citep{1980PhR....67..183K,1991PhLB..265..258V,1997PhRvD..55.4582S,2013PhRvD..87h3007K}, excitation of charged scalar fields during inflation ($10^{-30}\,\mathrm{s}$) \citep{1981PhRvD..23..347G,1982PhRvL..49.1110G,1988PhRvD..37.2743T}, and baryogenesis ($10^{-10}\,\mathrm{s}$) that leads to asymmetry between baryons and antibaryons \citep{1969Natur.223..938G,1970ApJ...160..451M,2010PhRvD..82b3008N}. We recommend the reader to the introduction of \cite{2009PhRvD..80b3013M} and the review by \cite{2012SSRv..166...37W} for a discussion of additional candidates. Regardless of the fields' origin, their importance on large scales at $15 \leq z \leq 30$ continues to be a mystery \citep{2014MNRAS.445.3706V}. It is believed that an efficient Biermann Battery mechanism during these redshifts can amplify the initial seed field \citep{1950ZNatA...5...65B,2008ApJ...688L..57X,2011ApJ...741...93D}; there are a number of other possible amplification mechanisms, including vorticity present in the primordial plasma during the radiation era  \citep{1970MNRAS.147..279H,1978MNRAS.184..843B} or an inverse energy cascade where magnetic energy is transferred from small to large scales \citep{1996PhRvD..54.1291B,2000PhRvD..62j3008F,2001PhRvE..64e6405C}.

Given the disagreement even over field amplification mechanisms, it is not surprising that there is a great deal of uncertainty about the field strength, which is exponentially sensitive to the source and physical parameters used \citep{2005PhR...417....1B}. Published estimates of the primordial magnetic field on correlation lengths of $50\,\mathrm{kpc}$ or more range from $10^{-34}\,\mathrm{G} - 10^{-9}\,\mathrm{G}$ \citep{2006MNRAS.372.1060T,2012PhRvD..86f3003K,2013ApJ...770...47K,2018CQGra..35o4001H,2018SSRv..214..122D}. For example, \cite{2006Sci...311..827I} use the cosmological power spectrum of magnetic fields to predict field strengths at the onset of primordial star formation of $10^{-18}\,\mathrm{G}$ on $1\,\mathrm{Mpc}$ and $10^{-14}\,\mathrm{G}$ on $10\,\mathrm{kpc}$ scales, respectively. Similarly, \cite{2004PhRvD..70l3003B} propose that primordial magnetic fields of strength $\sim 10^{-11}\,\mathrm{G}$ exist in galaxy clusters with correlation lengths of a few $\mathrm{kpc}$. 
\citet{2004ApJ...609..467M,2007PASJ...59..787M}, \citet{2015ApJ...801...13S}, and \citet{2018MNRAS.475.3331H} show that primordial molecular clouds are closer than modern-day ones to the limit of ideal magnetohydrodynamics (MHD) due to higher ionization fraction in the early Universe (see, however, \citealt{2019MNRAS.488.1846N}). No matter how the field is generated, it can be quickly amplified within a collapsing molecular cloud as a result of flux-freezing. Fields can be even further amplified during the collapse by the small-scale dynamo, which converts the kinetic energy of collapse-driven motions to magnetic energy \citep{2005PhR...417....1B}. Simulations show that this mechanism rapidly increases the magnetic energy density to a significant fraction (up to $\sim 50$ per cent) of the kinetic energy density \citep{2011PhRvL.107k4504F,2014ApJ...784...94S,2015PhRvE..92b3010S,2016JPlPh..82f5301F}. All the necessary conditions for the existence of a small-scale dynamo are fulfilled in the early Universe prior to primordial star formation \citep{2010ApJ...721L.134S,2012MNRAS.423.3148S,2012ApJ...754...99S,2014PhRvD..89j3001W}, as cosmological simulations directly predict the existence of turbulence at the onset of primordial cloud collapse \citep{2007ApJ...665..899W,2008ApJ...682..745W,2008MNRAS.387.1021G}.

While these studies strongly favour the presence of a dynamically-significant magnetic field during the formation of the first stars, there have been only limited explorations of how this affects the star formation process, and in particular the IMF of the first stars. Some of the first 3D nested-grid MHD simulations were performed by \cite{2006ApJ...647L...1M,2008ApJ...677..813M,2008ApJ...685..690M}, who  find that strong, ordered magnetic fields lead to the formation of jets and outflows. \cite{2012ApJ...745..154T} and \cite{2013MNRAS.432..668L} simulate the effects of magnetic fields on Population III star formation starting from cosmological structure conditions. They show that the field can be quickly amplified by the action of the small-scale dynamo if the resolution is sufficient to resolve the turbulent motions of the gas. However, they are unable to study the primordial IMF because their numerical techniques do not allow them to run past the formation of the first collapsed object. \cite{2013MNRAS.435.3283M} perform MHD simulations of primordial clouds with varying magnetic field strengths and find that fragmentation scales with the inverse of the field strength, with stronger fields resulting in the formation of a single, massive star. However, their resolution is insufficient to capture dynamo effects, and their numerical method precludes them from considering turbulent initial conditions. Thus, it becomes clear that the effects of magnetic fields on Population III star formation largely remain unexplored \citep{2013RPPh...76k2901B,2019ffbh.book...67K,2020SSRv..216...48H}.

This work adds on to earlier investigations of magnetic fields during primordial star formation in two straightforward ways: (1.) we study the effects of turbulent, non-uniform magnetic field structures, which has not been investigated in previous simulations. Once the collapse sets in, the turbulence being driven by gravity in the center of the minihalo will quickly convert even an originally uniform field to a tangled one with a randomly oriented geometry \citep{2004ApJ...612..276S,2005PhR...417....1B}, and (2..) we carry out 200 simulations with different realisations of the initial turbulent velocity and magnetic field distribution, so that we can sample enough stars to conduct a meaningful analysis of the resulting mass distribution. As we show later (see also, \citealt{2020MNRAS.494.1871W}), the amount of fragmentation highly varies within different realizations of the same field strength, thus, we cannot draw statistically meaningful conclusions unless we collect enough statistics to overcome the effects of stochasticity \citep{2013ApJ...776...48H,2016MNRAS.455.1438Y}.

This paper is organised as follows: \autoref{s:method} discusses how we setup the simulation and the turbulent magnetic field strengths we use, \autoref{s:results} presents our results, with a discussion in \autoref{s:discussion}. We summarise our findings and conclusions in \autoref{s:conclusions}.

\section{Numerical simulation methods}
\label{s:method}
\subsection{Simulation Setup}
\label{s:simulation_setup}
We utilize the adaptive mesh refinement (AMR) code FLASH \citep{2000ApJS..131..273F,2008ASPC..385..145D,Dubey_2013} for our MHD simulations. We largely adopt the simulation setup of \cite{2019MNRAS.490..513S}, with important additions including magnetic fields and deuterium chemistry. We use a tree-based solver to solve the Poisson equation for self-gravity \citep{2018MNRAS.475.3393W}, the five-wave approximate Riemann solver HLL5R to solve the MHD equations \citep{Bouchut2007,Bouchut2010} adopted for FLASH by \cite{2009JCoPh.228.8609W,2011JCoPh.230.3331W}, and sink particles as a proxy for stellar particles that form during collapse \citep{2010ApJ...713..269F,2011IAUS..270..425F,2014ApJ...797L..19F}. The sink particle module has been shown to work well for MHD simulations in FLASH because the grid-based implementation can ensure the magnetic field geometry remains intact during sink particle creation \citep{2011MNRAS.417.1054S,2011IAUS..270..291D}; further, the checks required for a sink to form contain appropriate contributions from magnetic pressure (for Jeans mass) and energy (for virial checks).

\autoref{tab:inicond} lists the initial conditions we utilize in this work. We start with a primordial cloud core of mass $1000\,M_{\odot}$ and a radius of $1\,\mathrm{pc}$. The size of the computational box we use is $L = 2.4\,\mathrm{pc}$. These initial conditions are in very good agreement with the overdense regions observed in dark matter mini-haloes in cosmological simulations that form around $z \sim 30$ \citep{2015MNRAS.448..568H}. Similarly, taking inspiration from cosmological simulations and MHD simulations of contemporary star formation, we initialise an angular velocity in the core to initiate a solid-body rotation around the $\hat{z}$ axis, such that the rotational energy is 3 per cent of the gravitational energy at the initial stage. Our base grid consists of $8^3$ cells on top of which we add up to 14 levels of refinement; the maximum effective resolution of the simulations thus reaches $65536^3$ cells. The maximum physical resolution and number density the simulations reach is given by the minimum cell size of $dx = 7.6\,\mathrm{au}$, and maximum gas density of $n \sim 10^{13}\,\mathrm{cm^{-3}}$ (equivalent to a mass density $\rho \sim 10^{-11}\,\mathrm{g\,cm^{-3}}$), respectively.

\subsection{Primordial chemistry}
\label{s:krome}
Since chemical evolution timescales in primordial clouds are comparable to collapse times \citep{1998ApJ...508..141O}, it is necessary to solve the chemical network associated with them in order to self-consistently compute the temperature as a function of the density during the collapse. We use the chemistry package KROME \citep{2014MNRAS.439.2386G}, which is designed to be incorporated in astrophysical simulations where treating the chemistry and hydrodynamics together is a crucial requirement \citep{2013MNRAS.431.1659G}. Specifically, we utilize the primordial chemistry network in KROME, which includes the following species: H, D, H$_2$, HD, H$^+$, D$^+$, H$^-$, D$^-$, He, He$^+$, HD$^+$, He$^{++}$, H$^{+}_{2}$ and $\mathrm{e^-}$. We run a 1D primordial cloud collapse model in KROME to generate initial mass fractions, core temperature and density that we supply as inputs to our 3D simulations\footnote{In the primordial chemistry network in KROME, we have adjusted the break-point in temperature of the reaction rate coefficient for the reaction \hii + D $\to$ HD + H (adopted from reaction IX17, Appendix A of \citealt{2009MNRAS.393..911G}) for maintaining numerical stability at high densities where the flux of this reaction is high. This adjustment ensures that the rate coefficient changes smoothly as a function of temperature, and is consistent with the experimental estimates originally provided by \cite{2003PhRvL..91f3201M} for this reaction.}. The mass fractions of the ionized species returned by KROME are scaled to ensure charge neutrality is maintained. We also follow an accurate calculation of the \hii adiabatic index (\gammah) implemented by \cite{2019MNRAS.490..513S}, which takes into account that, in the temperature range crucial for the formation of the first stars, quantum effects for \hii are non-negligible, and thus the gas is not well-approximated as a classical diatomic gas with $\gamma_{\rm H_2}=7/5$. For all other species, we assume their adiabatic indices to be $5/3$, apart from the remaining diatomic species (HD, HD$^+$ and H$^+_2$) for which we assume it to be $7/5$. The net adiabatic index of the gas is then given by the mass-weighted average of adiabatic indices of all species. Further, we implement a strict temperature floor given by the CMB temperature at our assumed redshift $z=30$. 

KROME also contains inbuilt functions to estimate the net heating and cooling contributed by the chemistry during the collapse. Specifically, we include cooling due to \hii, Lyman-$\alpha$ cooling, collisionally-induced emission (CIE) cooling, cooling due to Compton scattering of CMB photons, cooling due to HD, and cooling and heating due to chemical reactions that can be either exothermic and/or endothermic. At high densities, we use the opacity correction given by \cite{2004MNRAS.348.1019R} in the cooling function for \hii. We refer the reader to \citet[their Section 3.2]{2019MNRAS.490..513S} for a discussion of the caveats associated with the implementation of these processes, and point out that the density and temperature spaces covered by our simulations are not significantly affected by the approximations used in KROME, at least prior to the onset of radiation feedback which we do not consider in this work.

\subsection{Turbulence}
\label{s:turb_bfields}
Once an overdense region starts forming in the center of the mini-halo, it creates a potential well that pulls the baryons inwards and causes an infall of the gas. The dynamics of such a collapsing system naturally lead to the production of turbulence \citep{2008MNRAS.387.1021G}, which is a crucial ingredient for star formation. Turbulence can also be generated in the early Universe due to streaming velocities of the baryons with respect to the dark matter particles as per the $\lambda$CDM model \citep{2010PhRvD..82h3520T,2011ApJ...736..147G,2011MNRAS.412L..40M}, and by primordial magnetic fields through density perturbations \citep{1996ApJ...468...28K}. Further, turbulence can be sustained and driven by gravity that creates compressive as well as solenoidal flows of gas during infall \citep{2011ApJ...731...62F}, increasing the density and temperature. Taking this into account, we initialize our simulations with different random turbulent fields such that the root-mean-square (rms) Mach number ($\mathcal{M}_{\mathrm{rms}}$) is 1, \textit{i.e.,} the initial velocity fluctuations equal the local sound speed. The initial velocity power spectrum goes as $P_{\mathrm{v}} \sim k^{-1.8}$ from wave numbers $k/(2\pi/L) = 2-20$ where $L$ is the length of the cubic computational domain (e.g., \citealt{2013MNRAS.436.1245F,2019MNRAS.485.5532G,2019MNRAS.486.3647K}). We select the power-law scaling to be between the Kolmogorov ($k^{-5/3}$, \citealt{1941DoSSR..32...16K}) and Burgers ($k^{-2}$, \citealt{BURGERS1948171}) turbulence. Since both these kinds of turbulence are primarily driven on large scales, the results do not sensitively depend on the initial spectrum \citep{2011ApJ...731...62F}. The turbulence is driven by mixed modes comprised of solenoidal as well as compressive motions \citep{2010A&A...512A..81F,2012ApJ...761..156F}.

We note that the turbulence here is artificially driven by setting up an initially turbulent velocity field as described above, but that there is no subsequent mechanism to continue driving the turbulence. In reality, the kinetic energy on the largest model scales would be provided by even larger scale interactions in the cosmic web, thus continuously driving the turbulence on model scales. In practice, this makes little difference for our simulations, because the duration of our runs is relatively small compared to the turnover time of the largest turbulent structures, which is comparable to the free-fall time evaluated at the initial mean density. Thus, little turbulent decay occurs within the duration of our runs. However, we are currently in the process of constructing cosmological simulations that will be used as initial conditions for zoom-in simulations, in a forthcoming study. This will allow runs with significantly longer durations. A possible alternative to cosmological initial conditions would be to continuously drive the turbulence, but in a much larger computational domain containing multiple collapsing cores. However, this is not as realistic as cosmological initial conditions, and is computationally much more expensive, because if there are multiple collapsing cores present, then each must advance in time using the most stringent time step constraint that applies to any of them. This constraint would effectively mean that we could not follow the simulation much past the point where the first few cores collapsed. 

Additionally, in line with the arguments made by several previous works about the necessity of sufficiently resolving Jeans length to capture the effects of the small-scale dynamo \citep{2010ApJ...721L.134S,2011ApJ...731...62F,2012ApJ...745..154T,2013ApJ...772L...3L}, we set our AMR refinement condition to maintain at least 32 cells per Jeans length. Note that in AMR MHD simulations, it is also common to refine the grid based on derivatives of the velocity field (such as vorticity or divergence), as they can better capture the turbulent energy cascade responsible for shocks, especially in cases where the evolution of low-density regions is crucial \citep{2006ApJ...638L..25K,2008MNRAS.388.1079I,2009JPhCS.180a2020K,2009A&A...504...33V,2009A&A...494..127S,2014MNRAS.445.3706V,2017MNRAS.469.3641I}. However, \citet{2007ApJ...665..416K} and \citet{2009A&A...494..127S} show that such refinement criteria can be more computationally expensive than Jeans refinement, because they are extremely sensitive to the conditions at the shock front that can change by a large margin, thus reducing the simulation timestep and making long-duration runs difficult to accomplish. Furthermore, using such criteria would lead us to refine large portions of the computational domain that contain mostly low-density material that will never be accreted or interact with the dense regions that are the focus of this study, highly increasing the computational cost for no gain in accuracy in the regions about which we care. We can capture the regions in which we are interested sufficiently well by resolving the Jeans length with 32 cells, including the turbulent energy content on the Jeans scale \citep{2011ApJ...731...62F}.

\subsection{Magnetic Fields}
\label{s:turb_bfields2}
We use four model cases with different magnetic field strengths to study the role of magnetic fields in setting fragmentation early on during the collapse of primordial clouds. The four cases, as we show in \autoref{tab:inicond}, are named $B_0 - B_3$, and have initial rms field strengths of $0\,\mathrm{G}$, $1\,\mathrm{fG}$, $9\,\mathrm{\mu G}$ and $28\,\mathrm{\mu G}$, respectively. $B_0$ is an ideal case with no magnetic field strength that acts as a control simulation. Our motivation for case $B_1$ is to test an unlikely condition where an initial seed field has not already been amplified due to the small-scale dynamo at the onset of collapse, and the field strength is close to the pre-dynamo values discussed in \autoref{s:intro}; such a case seems unlikely because if the small-scale dynamo is present, it will very quickly amplify any weak seed magnetic field \citep{2011ApJ...731...62F,2016JPlPh..82f5301F,2016RPPh...79g6901S}, even before the presence of a protostellar disc \citep{2010A&A...522A.115S}. In fact, saturation of the field due to the small-scale dynamo is also observed very early on ($n \sim 10^{5}\,\mathrm{cm^{-3}}$) even when non-ideal MHD effects like ambipolar and Ohmic diffusion are considered \citep{2012ApJ...754...99S}. 

The cases $B_2$ and $B_3$ demand a more qualitative as well quantitative reasoning. Even though simulations are now able to resolve the action of the small-scale dynamo by efficiently resolving the Jeans length, they have not yet reached convergence. In other words, the higher the number of cells per Jeans length are used, the more the amplification of the magnetic field during the formation of the first stars is observed \citep{2010ApJ...721L.134S,2011ApJ...731...62F,2012ApJ...745..154T,2013ApJ...772L...3L}. This is because the simulations can only reach kinematic Reynolds numbers (the ratio of flow scale to viscous dissipation scale) of up to $10^{2-4}$ \citep{2009JPhCS.180a2020K,2011MmSAI..82..588J}, in best possible cases up to $10^5$ \citep{2013MNRAS.436.1245F,2020IAUS..345...43F}, whereas star-forming regions and the ISM in the Universe typically have Reynolds numbers of the order of $10^7$ \citep{2011ApJ...737...13K,2014PhR...539...49K}. Since the current MHD simulations are not able to reach convergence, they cannot reliably show the limit in which the dynamo action saturates. Dynamo saturation occurs when the back reaction of the magnetic field on the gas due to the Lorentz force causes the peak of the magnetic energy spectrum to shift to the largest possible spatial scales. This is thought to occur when the diffusivity equals the growth rate of magnetic fields \citep{1999PhRvL..83.2957S,2002NJPh....4...84S,2015PhRvE..92b3010S}.  

Nevertheless, theoretical developments as well as isothermal MHD simulations of turbulence in a box \citep{2004PhRvE..70a6308H,2011ApJ...731...62F,2014ApJ...797L..19F,2015PhRvE..92b3010S} predict that the saturation rate of magnetic energy produced by the turbulent dynamo can be anywhere between a fraction of a per cent to a few tens of per cent of the turbulent kinetic energy, depending on the turbulent Mach number, Reynolds number of the flow and magnetic Prandtl number (ratio of magnetic to hydrodynamic Reynolds number). Given that the magnetic Prandtl numbers in the early Universe were high \citep{1995stf..book.....C,1999ARA&A..37...37K,2012ApJ...754...99S}, we expect the dynamo to have saturated at a few per cent of the turbulent kinetic energy at our chosen initial sonic Mach number (see Figure 3 of \citealt{2011ApJ...731...62F}, Figure 2 of \citealt{2014ApJ...797L..19F} and Figure 4 of \citealt{2016JPlPh..82f5301F}). It has also been shown that the field will saturate very quickly if strong accretion shocks are present \citep{2014MNRAS.440.1551L}. Hence, we initialize cases $B_2$ and $B_3$ with magnetic field strengths such that the initial magnetic energy is 1 and 10 per cent, respectively, of the total turbulent kinetic energy in the system. This gives $B_2 = 9\,\mathrm{\mu G}$ and $B_3 = 28\,\mathrm{\mu G}$, respectively. 

The associated initial magnetic power spectrum goes as $k^{3/2}$ over a wide range of wave numbers ($2 \leq k \leq 20$) in the simulation box, the so-called Kazantsev spectrum resulting from turbulent dynamo amplification \citep{1968JETP...26.1031K,1985ZhETF..88..487K,2014ApJ...791L..34B}. There is no well-defined orientation of the field with respect to the rotation axis since we work with non-ordered fields as expected from the action of the small-scale dynamo\footnote{There is still a possibility of an ordered component of the magnetic field that can be generated later on in the core via the $\alpha\omega$ type large-scale dynamo \citep{2016RPPh...79g6901S,2016A&A...585A.151L,2019arXiv191107898L}.}. We note that certain MHD simulations have shown that a strong magnetic field can alter the underlying velocity power spectrum \citep{2009ApJ...691.1092L,2012ApJ...750...13C,2020MNRAS.492..668B}. Thus, the velocity power spectrum could take a sightly different form for magnetized versus non-magnetized simulations; however, we ignore any such effects here as they would not significantly change the exponent of -1.8 appropriate for trans-sonic turbulence, which is a reasonable intermediate value between the Kolmogorov and Burgers exponents of -5/3 and -2, respectively.

\begin{table}
\centering
\caption{Initial conditions of the spherically homogeneous primordial cloud. The RMS magnetic field strength in cases $B_2$ and $B_3$ is also expressed as a fraction of the turbulent kinetic energy ($E_{\mathrm{turb,kin}}$).}
\label{tab:inicond}
\begin{tabular}{|lcr|}
\hline
Parameter & Symbol & Value\\
\hline
Cloud Mass & $M_{\mathrm{core}}$ & $1000\,M_{\odot}$\\
Cloud Radius & $R_{\mathrm{core}}$ & $1\,\mathrm{pc}$\\
Cloud Number Density & $n_{\mathrm{core}}$ & $9050\,\mathrm{cm^{-3}}$\\
Cloud Temperature & $T_{\mathrm{core}}$ & $ 265\,\mathrm{K}$\\
Rot. / Grav. Energy & $E_{\mathrm{rot}}/E_{\mathrm{grav}}$ & $0.03$\\
Mass Fraction of H & $x_{\mathrm{H}}$ & $0.7502$\\
Mass Fraction of D & $x_{\mathrm{D}}$ & $4.56 \times 10^{-5}$\\
Mass Fraction of \hii & $x_{\mathrm{H_2}}$ & $0.0006$\\
Mass Fraction of He & $x_{\mathrm{He}}$ & $0.2492$\\
Mass Fraction of HD & $x_{\mathrm{HD}}$ & $3.82 \times 10^{-8}$\\
Mass Fraction of e$^-$ & $x_{\mathrm{e^-}}$ & $4.72 \times 10^{-9}$\\
CMB Temperature at $z=30$ & $T_{\mathrm{CMB}}$ & $84.63\,\mathrm{K}$\\
Turbulence & $v_{\mathrm{rms}}$ & $1.84\,\mathrm{km\,s^{-1}}$\\
Sound Speed & $c_{\mathrm{s}}$ & $1.84\,\mathrm{km\,s^{-1}}$\\
RMS Magnetic Field & $B_0$ & $0$\\
 & $B_1$ & $1\,\mathrm{fG}$\\
 & $B_2$ & $9\,\mathrm{\mu G}$ ($0.01\,E_{\mathrm{turb,kin}}$)\\
 & $B_3$ & $28\,\mathrm{\mu G}$ ($0.10\, E_{\mathrm{turb,kin}}$)\\
 \hline
\end{tabular}
\end{table}

\section{Results}
\label{s:results}
We run a total of 200 simulations, 50 realizations each for the four different initial magnetic field strengths we use, as shown in \autoref{tab:inicond}. We only change the random seeds of the initial turbulence and magnetic field distributions between the different runs. The set of 50 random seeds is identical for each magnetic field strength, so, for example, run 1 for cases $B_0$ - $B_3$ has the same initial velocity field and the same magnetic field structure in all four cases; only the field strength differs. Given that star formation is well known to be a stochastic process (e.g., \citealt{1978ApJ...223..129G,1979ApJ...232..702S,2006MNRAS.367.1394K,2011ApJ...741L..26F}), such simulations are an ideal way to study the overall pattern and distribution of a sample since they collectively take into account the changes one would expect simply from stochasticity \citep{2020MNRAS.494.1871W}.

Similar to \citet{2019MNRAS.490..513S}, we parameterize the time for which we run our simulations by the star formation efficiency, defined as the ratio of the total mass of sink particles $M_\mathrm{sink}$ to the initial cloud mass $M_\mathrm{cloud}$, i.e., $\mathrm{SFE} = M_{\mathrm{sink}} / M_{\mathrm{cloud}}$. We stop our simulations when SFE = 5 per cent, since we do not include radiation feedback, which starts to become significant for primordial stars $\gtrsim 50$ $M_\odot$ \citep{2011Sci...334.1250H,2020ApJ...892L..14S}. This threshold is usually achieved between $500-5000\,\mathrm{yr}$ after the first sink particle is formed. All the runs collectively form a total of 1157 sink particles. 

\begin{figure*}
\includegraphics[width=\linewidth]{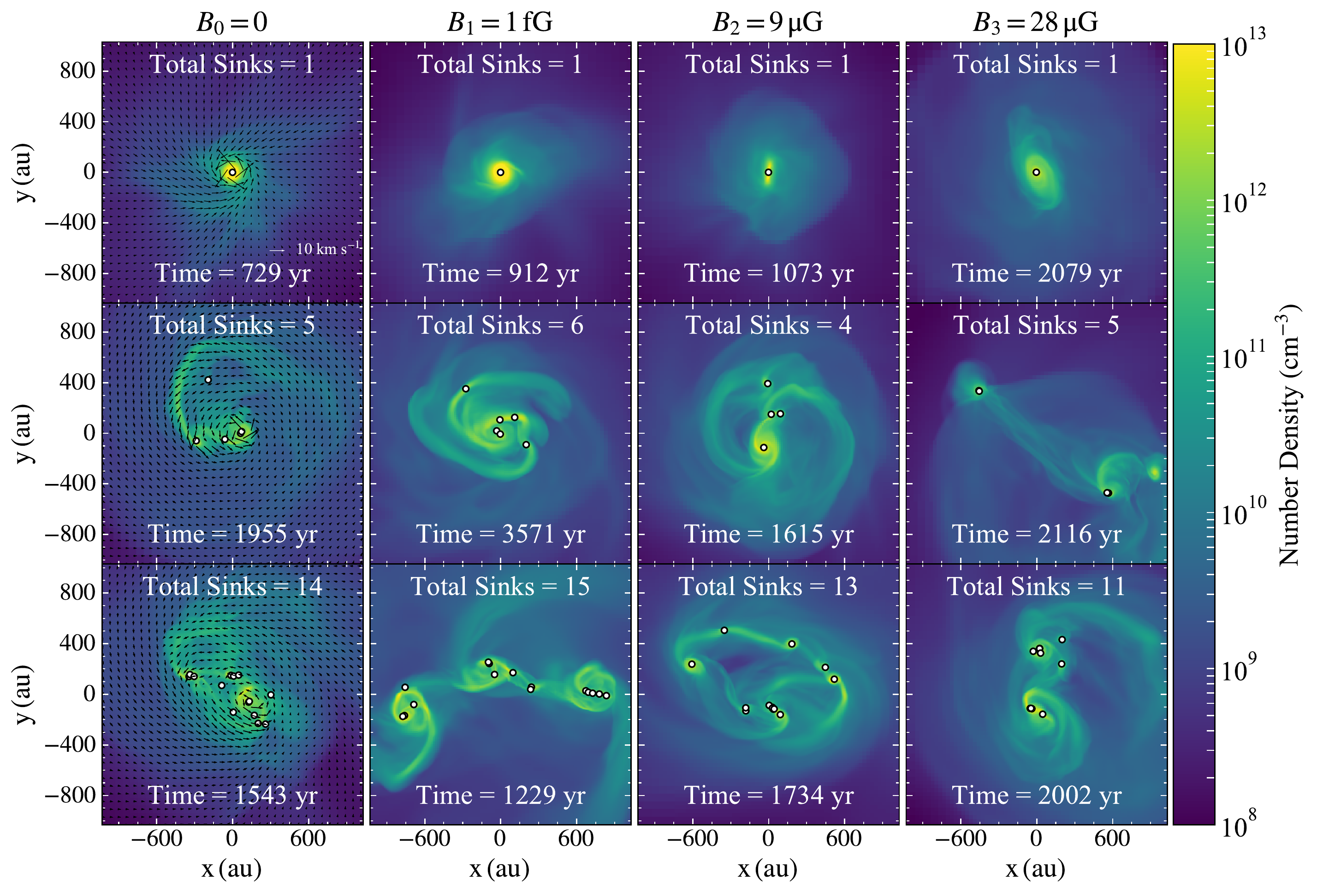}
\caption{Density-weighted projection maps (through the $\hat{z}$ axis) of the number density ($n$) for three randomly selected realizations from each of the four cases with different initial magnetic field strengths in each column. The random seed for the simulations shown in the first row is the same for all four cases, while it differs for the other two rows. These realizations depict the central $0.01\,\mathrm{pc}$ region and result in no, medium and high fragmentation, respectively (from top to bottom in every column). The maps correspond to a time when all the sink particles (white circles with black boundaries) have collectively accreted 5 per cent of the initial cloud mass (SFE = 5 per cent). Time in the panels is given as time since the formation of the first sink particle. The contours on the first column depict the velocity vectors of the gas in the $x-y$ plane.}
\label{fig:projnumdens}
\end{figure*}

\begin{figure*}
\includegraphics[width=\linewidth]{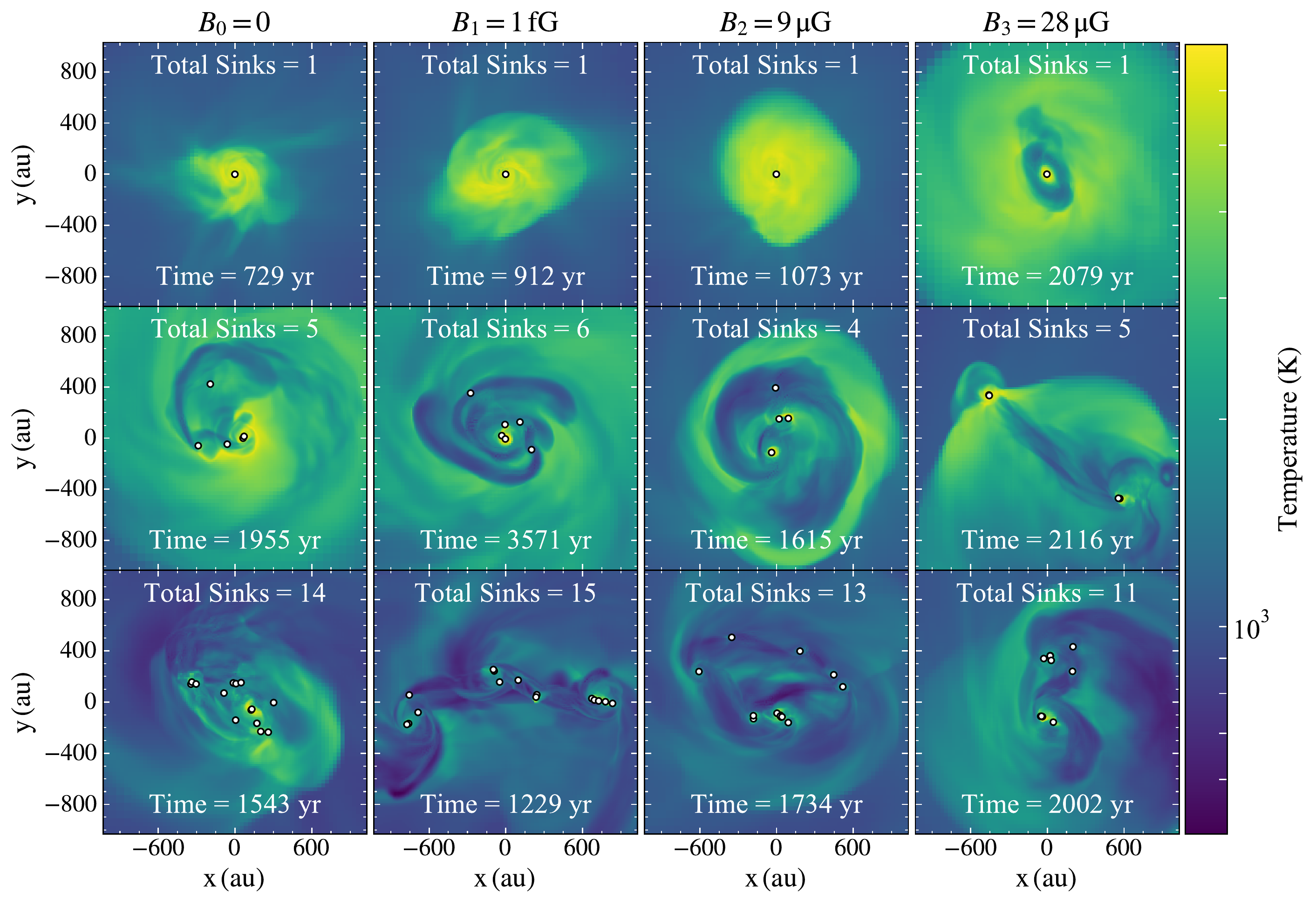}
\caption{Same as \autoref{fig:projnumdens} but for the density-weighted temperature. Both hot and cold accretion flows as well as spiral density patterns are noticeable. Cooler regions are highly molecular with \hii being the dominant species. Lyman-$\alpha$ cooling becomes effective at temperatures > $10^4\,\mathrm{K}$.}
\label{fig:projtemp}
\end{figure*}

\begin{figure*}
\includegraphics[width=\linewidth]{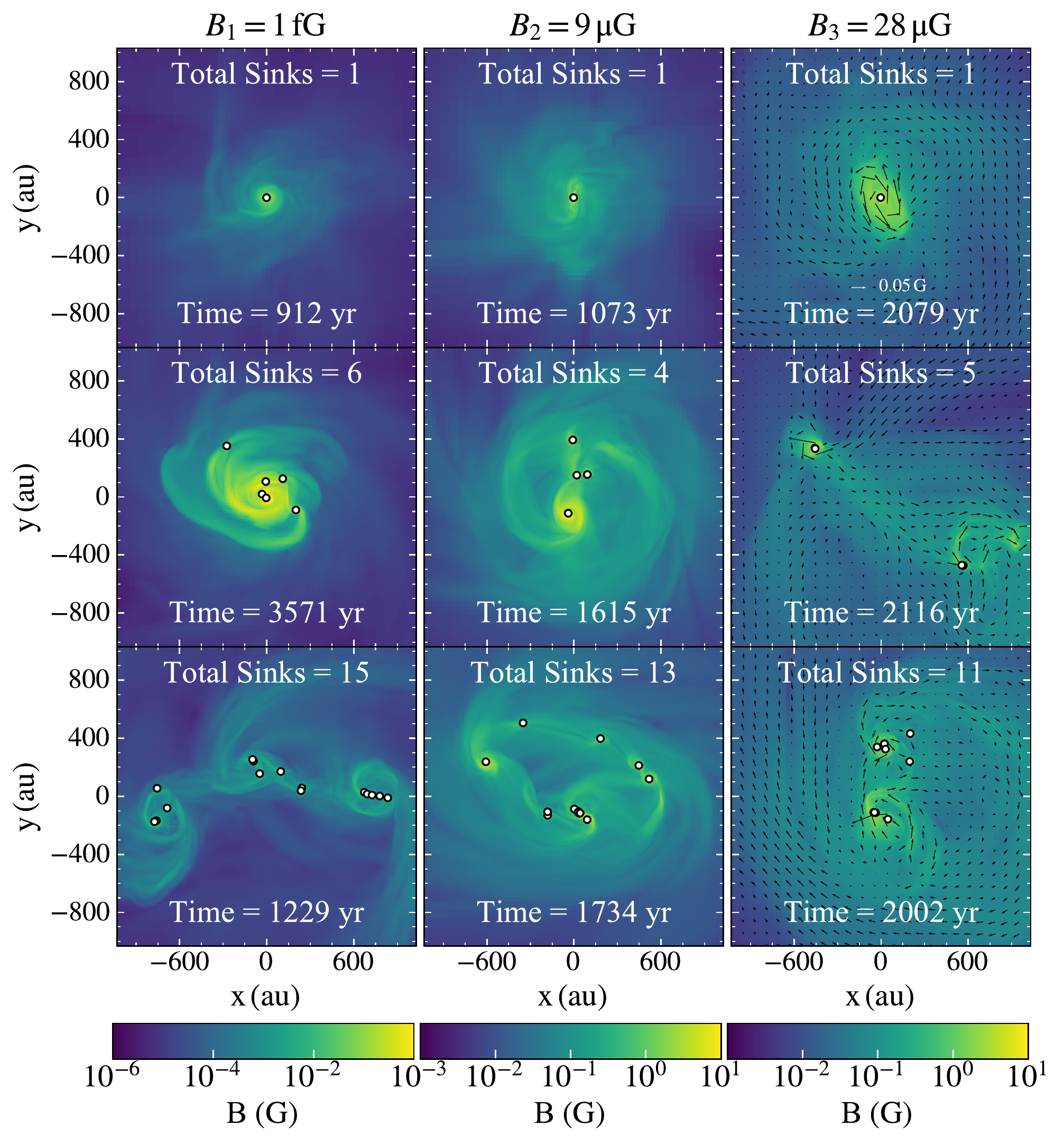}
\caption{Same as \autoref{fig:projnumdens} but for the density-weighted magnetic field strength in the three non-zero magnetic field cases. Arrows in the third panel mark the $xy$ components of magnetic field vectors. The length of all other vectors is a fraction (in $\log$) of the vector with the highest magnitude; for example, a vector half the length of the legend represents a field strength that is 10 times smaller.}
\label{fig:projbfield}
\end{figure*}

\begin{figure*}
\includegraphics[width=\linewidth]{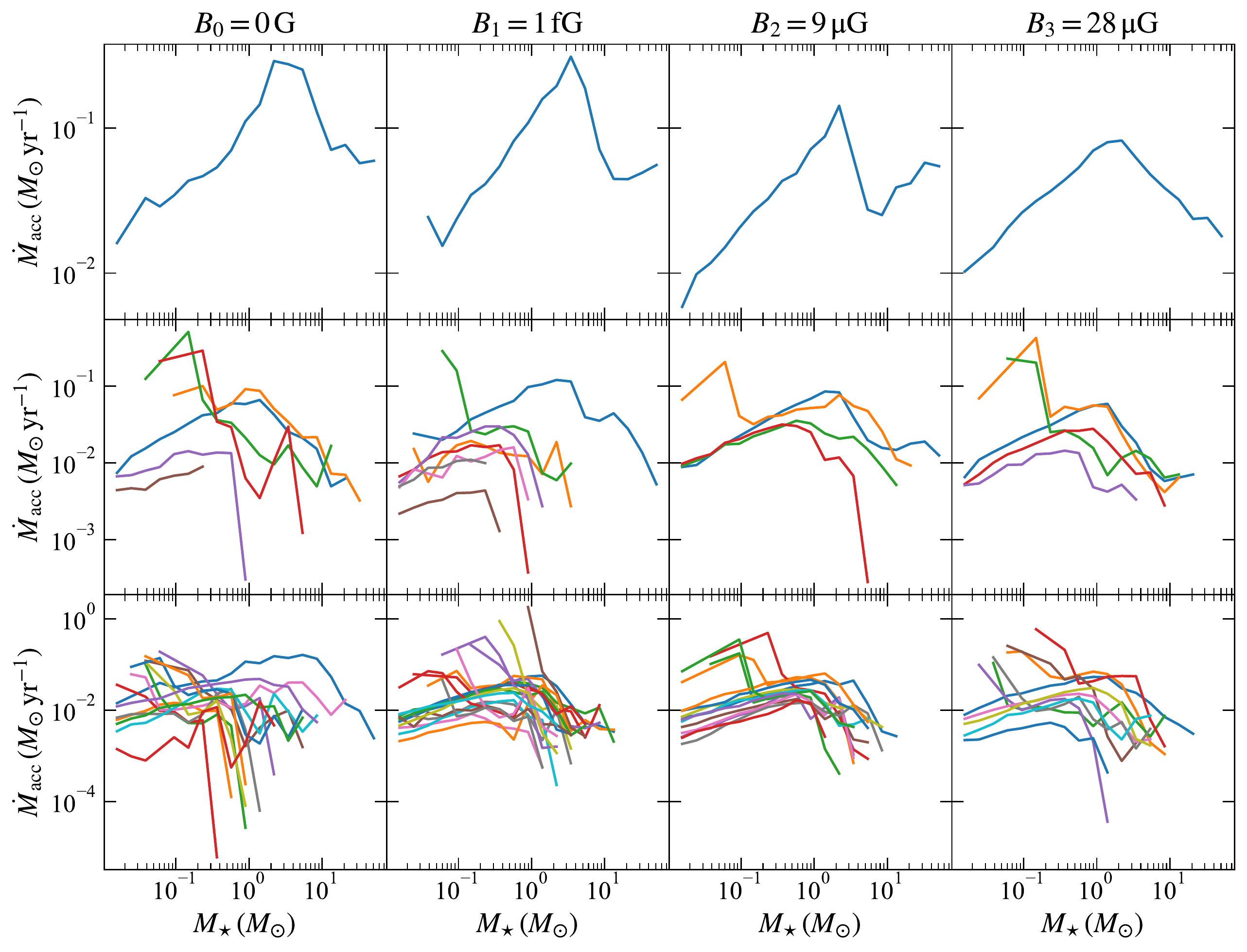}
\caption{Evolution of accretion rate against sink mass for the same simulations shown in \autoref{fig:projnumdens}, averaged over bins of sink mass. Each colored line represents an individual sink particle, with the blue depicting the first sink particle that forms in each case.}
\label{fig:accrates}
\end{figure*}

\begin{figure*}
\includegraphics[width=\linewidth]{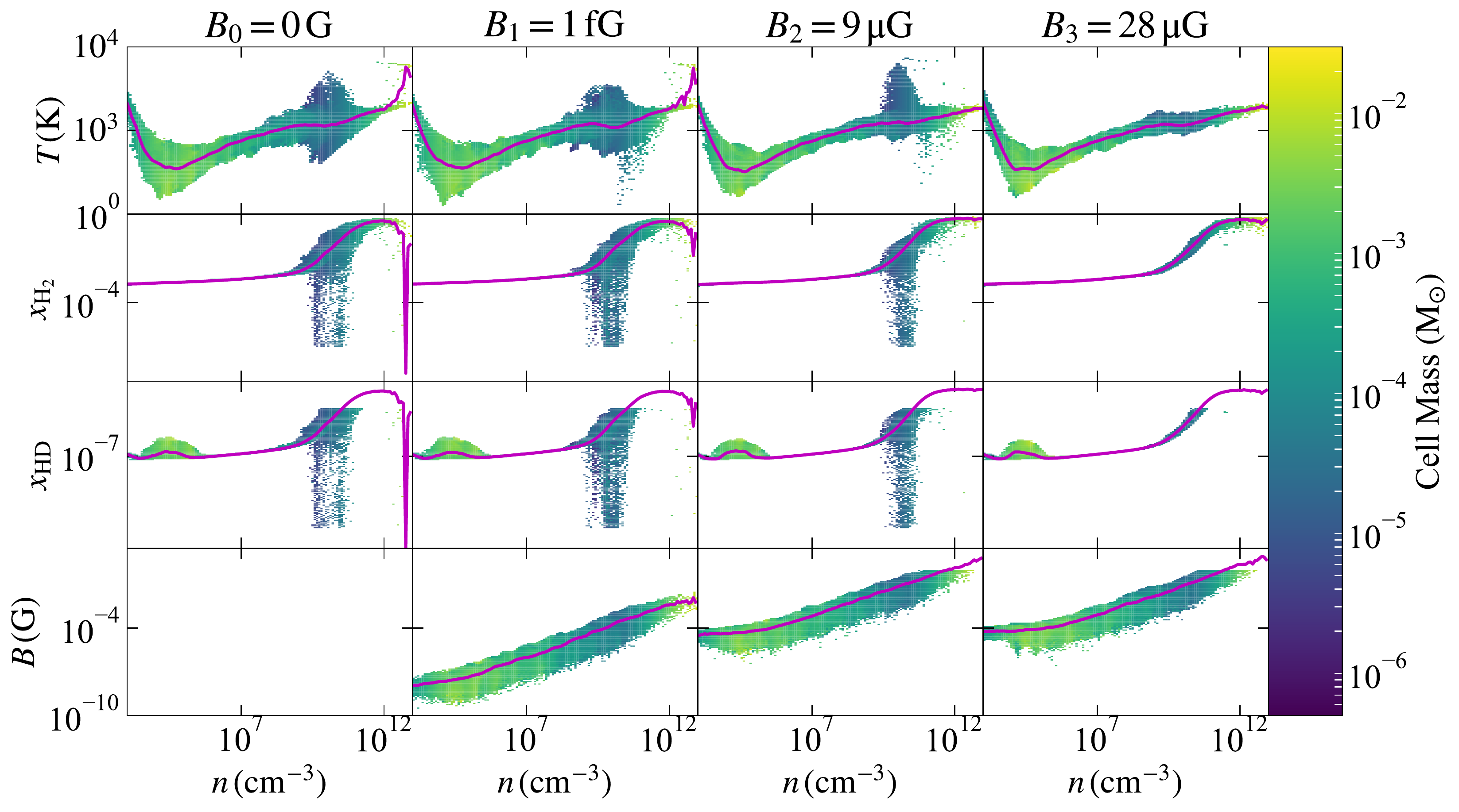}
\caption{Joint distributions of number density ($n$) as a function of temperature ($T$, first row), mass fraction of \hii ($x_\mathrm{H_2}$, second row), HD ($x_\mathrm{HD}$, third row), and the magnetic field strength ($B$, fourth row) of the gas in a randomly selected realization with the same random seed for all the four cases. These distributions represent a 0.5 pc sized region centered at the single sink particle that has just formed in the simulation. Magenta curves show the mean value of the quantity on the $y$-axis in bins of $n$.}
\label{fig:phaseplots}
\end{figure*}

\subsection{Morphology of discs and star systems}
\label{s:morhpsoutcome}
As the primordial cloud collapses, infall towards the center compresses the gas, leading to the creation of high-density peaks that ultimately form sink particles. In all cases, an accretion disc forms around the primary sink particle that may or may not fragment further to produce more sinks. We also find that the onset of collapse is delayed in cases where the magnetic field is strong. This is simply because magnetic pressure exerted on the cloud supports it against gravitational collapse \citep{2004MNRAS.347.1001H,2007MNRAS.377...77P}.

In \autoref{fig:projnumdens}, we show the projection of number density along the $z$ axis at SFE = 5 per cent for three pseudo-random realisations of each magnetic field strength that are selected to show no, few and high secondary fragmentation after the first sink particle has formed, respectively. Animated versions of this and related figures are available in the supplementary online material. It is straightforward to notice the diversity of systems formed in different cases only by changing the random seeds of turbulence or magnetic field. In runs where only one sink particle is formed, the accretion disc around it remains hot, inhibiting any further fragmentation, as seen in \autoref{fig:projtemp}. In cases where high fragmentation is observed, the discs are generally cooler. This is because the angular momentum transport causes the disc to spread out in radius, which allows the growth of density perturbations that can form multiple high-density peaks, which collapse to give rise to more sink particles \citep{1993MNRAS.264..798B}. In such cases, we find that sinks often tend to redistribute themselves to form clusters (at least for a short period of time), with an accretion disc associated with each cluster and large-scale high-density spiral patterns. They also result in the formation of numerous sub-solar and solar-type sinks, many of which remain bound to a massive ($M_{\star} > 20\,M_{\odot}$) primary \citep{2016MNRAS.462.1307S}. \autoref{fig:projnumdens} and \autoref{fig:projbfield} show the coupling between magnetic field and primordial gas that results due to flux-freezing.

In general, we observe reduced fragmentation as we increase the magnetic field strength. Random seeds which lead to the formation of only one sink particle in the $B_0$ case also form just one sink particle in all other cases. However, there are exceptions to this general trend, in the form of realisations where we observe more fragmentation in runs with strong fields. Fragmentation often also occurs in spiral density waves that develop due to gravitational instabilities and decreased local Jeans mass and sound speed \citep{2011MNRAS.417.1928F}, as can be seen from \autoref{fig:projtemp}. Runs where high fragmentation is observed often result in all sink particles being co-planar, as we see from projections along all the three axes.

\subsection{Evolution with time and gas density}
In \autoref{fig:accrates}, we plot the accretion rates of all sink particles averaged over bins of sink mass; the blue curve in each panel depicts the accretion rate of the first sink particle that is formed in the simulations. The accretion rates are generally in good agreement with similar studies (e.g., \citealt{2011ApJ...727..110C,2013ApJ...772L...3L,2014ApJ...781...60H,2014ApJ...785...73S,2018MNRAS.479..667R,2020MNRAS.494.1871W}). We also find that the accretion rates seldom drop below $10^{-4}\,M_{\odot}\,\mathrm{yr^{-1}}$ till SFE = 5 per cent where the effects of protostellar ultraviolet (UV) feedback becomes important \citep{2015MNRAS.449...77L}. The first row shows how magnetic fields affect accretion onto the sink particles by systematically lowering the peak as well as the overall accretion rate with time, similar to the findings of \cite{2007MNRAS.377...77P} for present-day star formation.

\autoref{fig:phaseplots} shows the evolution of temperature, mass fractions of \hii and HD, and the magnetic field strength as a function of number density just after the formation of the first sink particle (\textit{i.e.,} at SFE = 0) in randomly selected realizations from each case. The mean thermal evolution as shown in the first row is broadly in good agreement with the one-zone calculations of \cite{2005ApJ...626..627O}, and all other reported simulations of the first stars. The distributions of mass fractions of \hii and HD show a tighter correlation with their mass-weighted mean as the field strength increases. The dip in the temperature at low densities ($n \sim 10^5\,\mathrm{cm^{-3}}$) is a result of the onset of cooling due to the formation of HD during collapse at these densities \citep{2002ApJ...564...23B,2002ApJ...569..549N,2005ApJ...626..627O}. Even though the initial field strength for cases $B_2$ and $B_3$ differs by a factor of 3, the maximum field strength at the end of the simulation is similar. This might be due to the back reaction of the strong field on the density. We provide a more thorough analysis of the growth of magnetic field with density and its implications for dynamo action in a companion paper (P. Sharda et al., in preparation).

Another noteworthy feature of \autoref{fig:phaseplots} is that cells with the highest densities have lower mean temperatures in the strong-field cases ($B_2$ and $B_3$) than in the weak-field cases ($B_0$ and $B_1$); they also have correspondingly higher H$_2$ fractions, due to the lack of gas warm enough to induce collisional dissociation. This change occurs because, in the strong-field cases, shock compression that leads to temperature enhancements are limited by magnetic pressure. Our finding here is broadly consistent with that of \cite{2009ApJ...703.1096S}, who find that magnetic fields can change the thermal evolution of a collapsing primordial cloud. More importantly, it is also strong evidence that magnetic pressure plays a crucial role in reducing fragmentation: the more strongly-magnetised cases fragment less \textit{despite} having lower temperatures and thus less thermal support at high densities.

\begin{figure}
\includegraphics[width=\columnwidth]{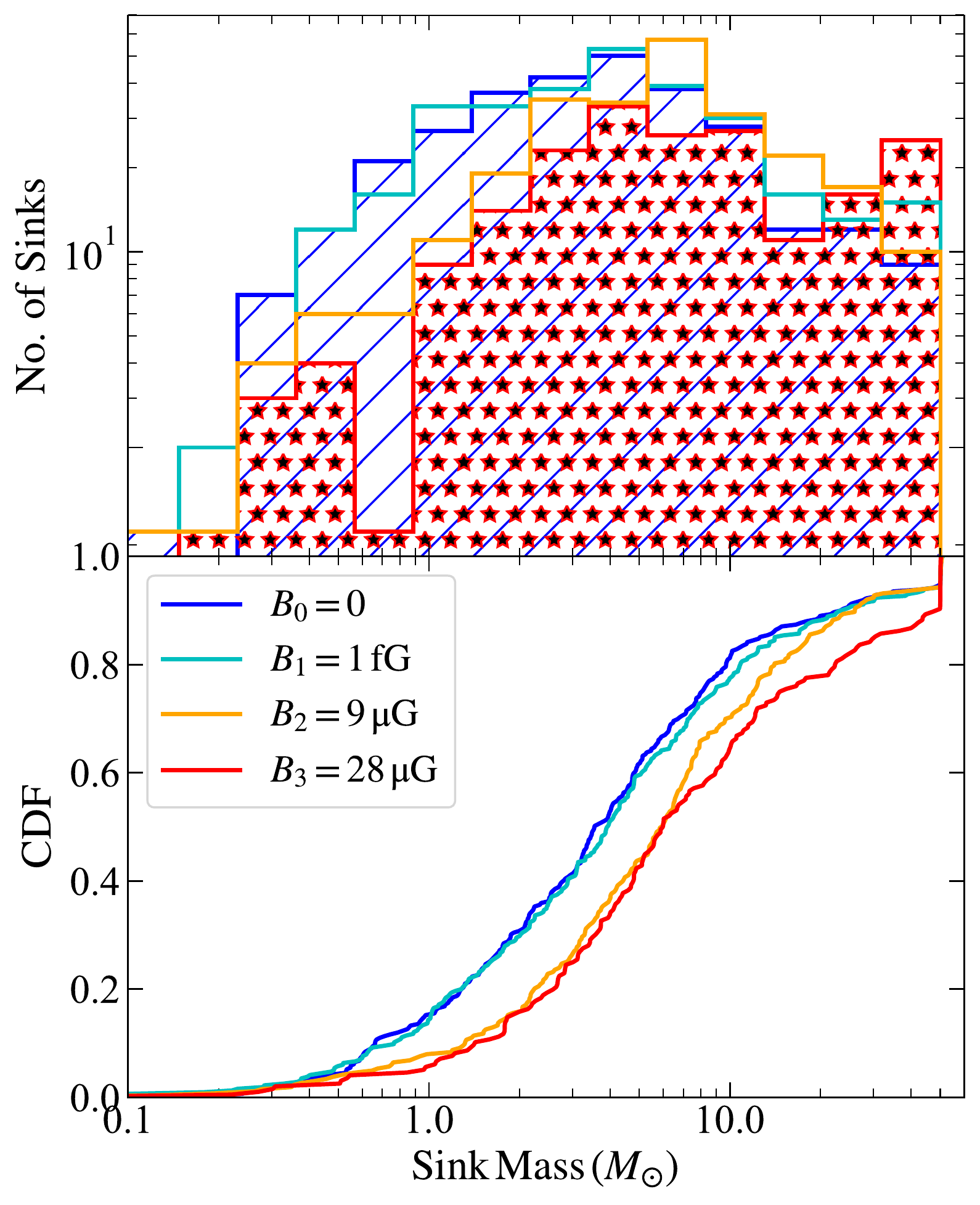}
\caption{Number of sink particles and their cumulative distribution function (CDF; \textit{bottom panel}) from all 200 simulations with different initial magnetic field strengths. The peak at $50\,M_{\odot}$ in the top panel and the corresponding jump in the CDF is due to runs where no fragmentation occurs, and our condition of stopping at $\mathrm{SFE}=5$ per cent therefore results in a single sink particle of mass $50\,M_{\odot}$.}
\label{fig:pdfcdf}
\end{figure}

\begin{table}
\centering
\caption{KS test $p$-values for different pairs of magnetic field strengths. If $p$-value < 0.01, there is less than a 99 per cent chance that the two sink mass distributions corresponding to the two magnetic field strengths are different.}
\label{tab:p-values}
\begin{tabular}{|l|c|c|c|r|}
\hline
$p$-value & B0 & B1 & B2 & B3\\
\hline
B0&1.0&0.87&$4.5\times10^{-5}$&$7.2\times10^{-5}$\\
\hline
B1&0.87&1.0&$5.2\times10^{-4}$&$4.5\times10^{-4}$\\
\hline
B2&$4.5\times10^{-5}$&$5.2\times10^{-4}$&1.0&0.25\\
\hline
B3&$7.2\times10^{-5}$&$4.5\times10^{-4}$&0.25&1.0\\
\hline
\end{tabular}
\end{table}

\begin{figure}
\includegraphics[width=\columnwidth]{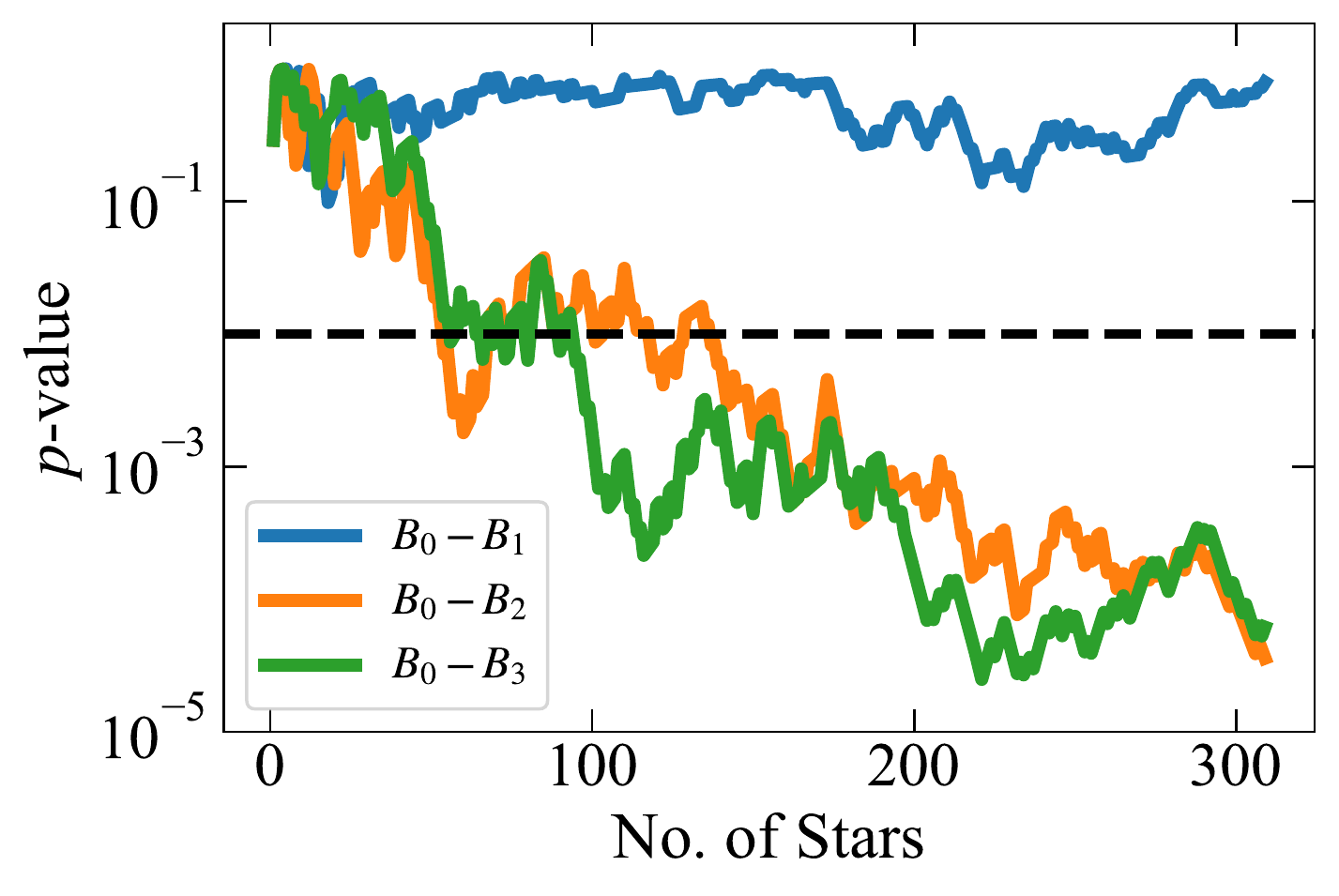}
\caption{Change in the $p$-value returned by comparing simulation $B_0$ to the other three cases ($B_1$ - $B_3$, as indicated in the legend) as a function of the mean number of stars being compared (\textit{i.e.,} a value of 100 means an average of 100 stars from each of the two runs). To construct this plot, we compute the $p$-value by comparing one realisation of $B_0$ to one realisation of $B_1$, then two realisations of each case, and so forth, and similar for $B_2$ - $B_4$. The dashed line denotes a $p$-value of 0.01, our adopted threshold for a significant detection.}
\label{fig:pvalue}
\end{figure}

\subsection{Initial mass function and multiplicity of the first stars}
\label{s:statsoutcome}

\subsubsection{Sink Mass Distribution}
\label{s:imf}
With 1157 sink particles formed across all simulations, we can perform a rigorous statistical analysis of the properties of the first stars that form under different initial magnetic field strengths. Figure \ref{fig:pdfcdf} shows the number of sinks and the cumulative distribution function (CDF) of the masses of sink particles formed in the four different magnetic field cases. The peak of the distribution of sink masses does not change appreciably between the four runs, however, a second peak at $50\,M_{\odot}$ becomes more and more prominent as the field strength increases. The latter simply represents the growing number of systems that remain single to SFE = 5 per cent, where we stop the simulations. Another prominent difference between the runs with zero/weak and strong magnetic fields is the smaller number of less massive sink particles (almost by a factor $\sim 2$ as seen in \autoref{fig:pdfcdf}), in runs where secondary fragmentation takes place; this is also clear from the separation in the CDFs for sink masses $>\,0.5\,M_{\odot}$. Thus, three important conclusions that we can draw from these observations are: (1) as the field strength increases, so does the chance of a first star evolving in isolation without any companions, (2) even if turbulent primordial clouds with strong initial magnetic fields do fragment, they tend to form fewer stars on average, and (3) strong magnetic fields suppress the formation of low-mass stars by a factor $\sim 2$ compared to cases where the field is weak or non-existent. 

To check whether the sink mass distributions resulting from simulations with different initial field strengths differ by a statistically-significant amount, we use the Kolmogorov-Smirnoff (KS) test, the output of which is a $p$-value with which we can rule out the null hypothesis that the sink masses in any given set of simulations were drawn from the same underlying distribution. Following \citet{2019MNRAS.490..513S}, we consider two distributions to be different if their corresponding KS test yields $p$-value $<0.01$, meaning we can rule out the null hypothesis at $>99$ per cent confidence. \autoref{tab:p-values} lists the $p$-values of the KS tests that we conduct between all possible pairs of simulations. The extremely low $p$-value for any pairs of $B_0$ or $B_1$ on one hand and $B_2$ or $B_3$ on the other, is strong evidence that the underlying sink mass distributions that result from the collapse of turbulent primordial clouds with weak/zero and strong magnetic fields are significantly different, with stronger fields yielding fewer, more massive fragments. This is consistent with expectations for contemporary star formation, where additional magnetic pressure increases the total (magnetic + thermal) Jeans mass and suppresses fragmentation \citep{2012ApJ...761..156F,2013MNRAS.430.1653H,2019FrASS...6....7K}.

Note that we derive this result based on 200 realisations in total. \autoref{fig:pvalue} demonstrates why such a large number of realisations is necessary, by showing how the $p$-value between the $B_0$ and other cases changes as we add more and more stars (from more and more independent simulations) to the sample. We see that detecting the difference between the mass distribution in case $B_0$ and those produced in cases $B_2$ or $B_3$ requires $\gtrsim 100$ stars (on average for each case, so $\gtrsim 200$ in total) for reliable detection. Smaller samples would be insufficient. This result reinforces our expectation, laid out in \autoref{s:intro}, that multiple realisations are necessary to draw strong conclusions about the characteristics of chaotic systems like turbulent star-forming molecular clouds.

\subsubsection{Multiplicity}
\label{s:multiplicity}
We next examine the number of singles, binaries, triples and quadruples formed in our simulations. The first step in our investigation is to classify stars by multiplicity. This is not a trivial task, since we cannot assume that all the stars formed in a single simulation constitute a bound system -- in some cases there are complex interactions between fragments in the disc that lead to one or more stars being ejected, such that they would likely end up single. To handle this issue, we classify every star formed in each simulation as single (S), or as belonging to a binary (B), triple (T) or quadruple (Q) system, following the algorithm given by \cite{2009MNRAS.392.1363B}. Briefly, this algorithm recursively finds the most bound stellar pair and replaces the constituent stars with a single star at their center of mass with a velocity equal to their center of mass velocity. The algorithm moves on to the next pair if the subsequent bound pair would lead to the formation of a quintuple, since such high-order multiples would almost certainly disintegrate dynamically. If no more bound pairs can be formed, the algorithm moves on to the next most bound pair among the remaining stars. The algorithm terminates if there are no more bound pairs, or if the only bound pairs remaining would, if combined, yield an aggregate of $>4$ stars. Once this state has been reached, the algorithm has classified every star in a given simulation into the type of system -- S, B, T, or Q -- to which it belongs.

\begin{figure}
\includegraphics[width=\columnwidth]{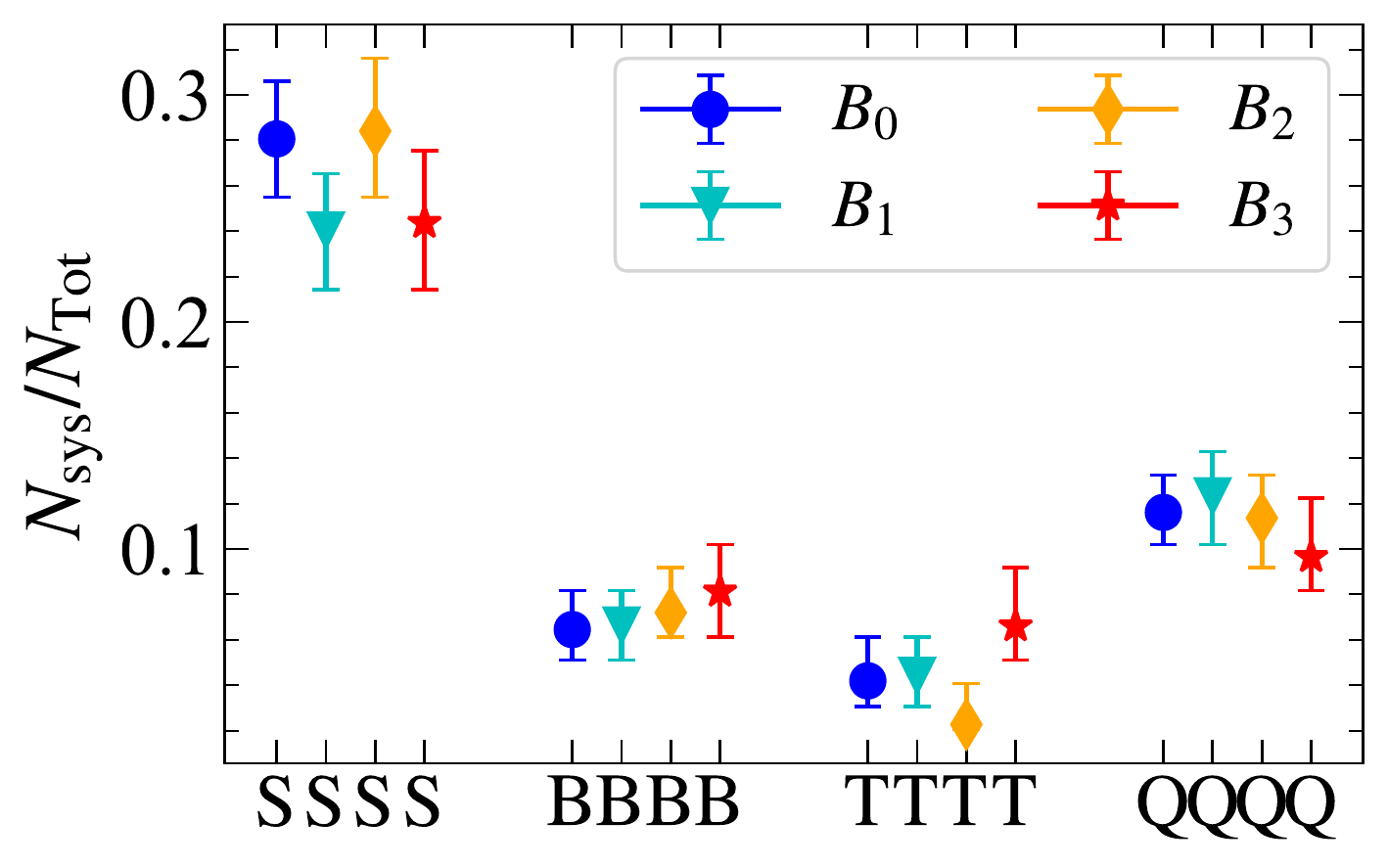}
\caption{Number of singles (S), binaries (B), triples (T) and quadruples (Q) formed in all the four main simulation cases (with different magnetic field strengths) divided by the total number of stars formed at the SFE = 5 per cent threshold, summed over all realizations. The error bars shown indicate the $16^{\mathrm{th}}$ to $84^{\mathrm{th}}$ percentile uncertainty range.}
\label{fig:sbtq_numbers}
\end{figure}

\autoref{fig:sbtq_numbers} plots the ratio of the number of singles, binaries, triples and quadruple systems to the total number of sink particles formed in each case. Following \cite{2012ApJ...754...71K}, we calculate the statistical uncertainty on these fractions by assuming that the number of stars we have in each case are random variates drawn from a binomial distribution for which the true probability that a randomly chosen star is single is $f$ (and similarly for all other multiplicities). We assume a flat prior on $f$ from [0,1]. Then, if the sample produced by our simulations constitutes exactly $M$ singles from a total of $N$ stars, the $16^{\mathrm{th}}$ percentile ($f_{16}$) on the posterior probability distribution for $f$ is then implicitly given by
\begin{equation}
\int_{0}^{f_{16}}  {N \choose M} f^{M}\,(1-f)^{N-M}\,df = 0.16\,,
\label{eq:f16}
\end{equation}
and the $84^{\mathrm{th}}$ percentile ($f_{84}$) is
\begin{equation}
\int_{f_{84}}^{1}  {N \choose M} f^{M}\,(1-f)^{N-M}\,df = 0.16\,.
\label{eq:f84}
\end{equation}
The median value $f_{50} = N/M$, not surprisingly. We see that, independent of magnetic field strength, almost one-third of all sink particles formed are single. Most interestingly, we find that the magnetic field strength has no effect on the multiplicity distribution. All four cases produce fractions of singles, binaries, triples, and quadruples that are identical within the statistical uncertainties, despite our large sample size.  

We can understand how to reconcile the apparent insensitivity of multiplicity to magnetic field strength with the clear dependence of the IMF on it by examining the mass functions broken down by stars that are classified into different multiplicity groups. To this end, we perform KS tests to check whether each pair of runs differs significantly for a particular multiplicity, for example, we ask whether the mass function for singles formed in case $B_0$ is statistically-distinguishable from the mass function for singles formed in cases $B_1$, $B_2$ or $B_3$. We provide the results of this analysis in \autoref{tab:table3} where we list the KS test $p$-values between the different pairs of multiples for different cases of initial magnetic field strength. This table provides us with important information on what drives the sink mass distributions with zero/weak fields to differ from those produced in simulations with strong fields. As we report in \autoref{s:imf}, the $p$-value between $B_0/B_1 - B_2/B_3$ cases is extremely low, indicating that the overall mass functions are statistically very different. From \autoref{tab:table3}, we see that the $p$-value for the distribution of binaries between these cases is very low, thus, it is clear that the binary population is strongly affected by the presence of magnetic fields. On the other hand, the $p$-value for triples and quadruples is high (except for quadruples of $B_1$ and $B_2$). Similarly, the $p$-value between the single sink distributions of $B_0/B_1 - B_2/B_3$ is also low. Thus, \autoref{tab:table3} indicates that the difference in the overall mass function between weak-field and strong-field cases arises primarily in cases where little or no fragmentation takes place, and the result is a single or binary. In cases where many fragments form, yielding a triple or higher, there is little difference.

\begin{figure}
\includegraphics[width=\columnwidth]{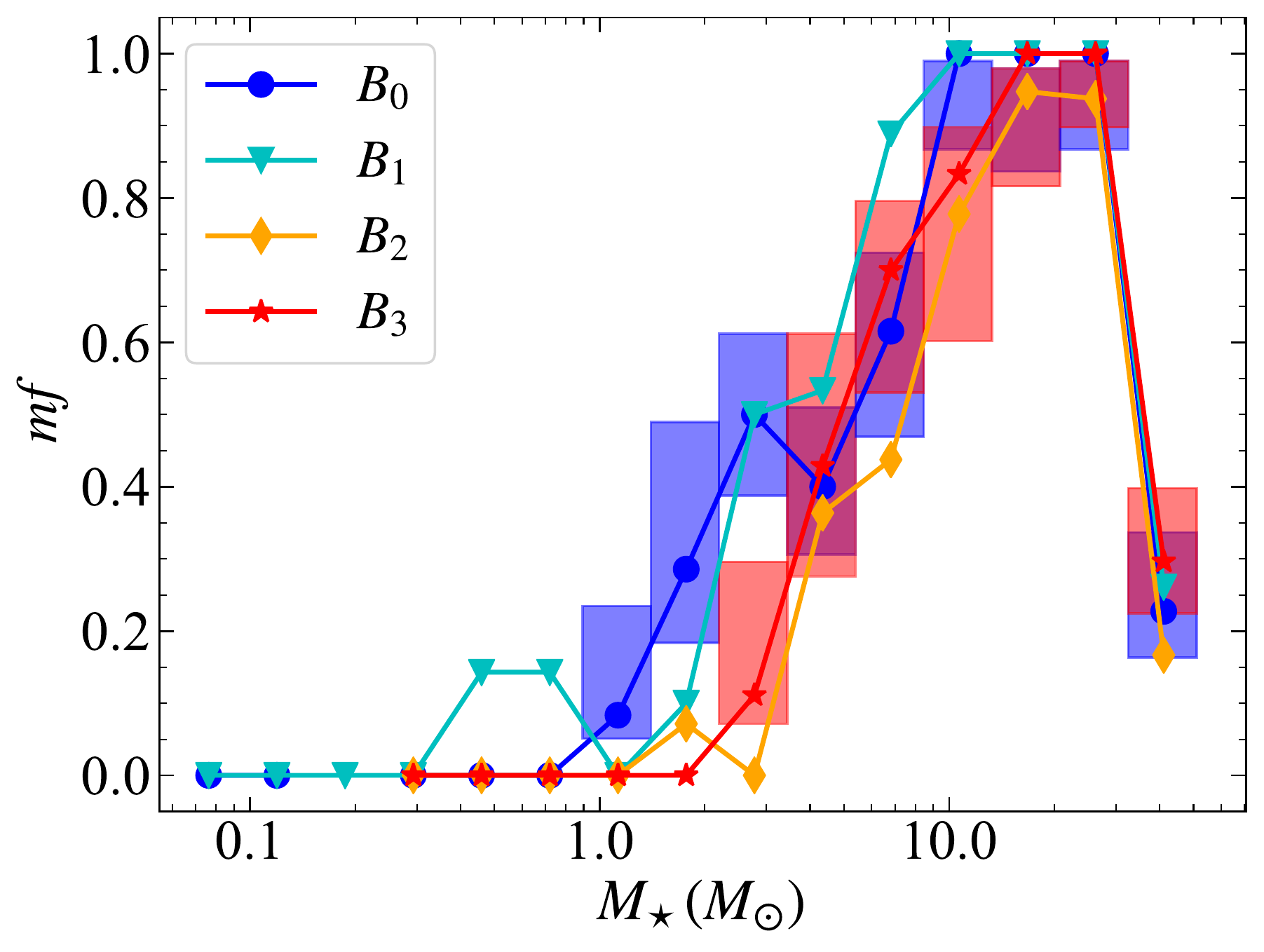}
\caption{Multiplicity fraction for each bin of primary mass, analogous to Figure 17 of \protect\cite{2012MNRAS.419.3115B} and Figure 14 of \protect\cite{2012ApJ...754...71K} for simulations of contemporary star formation. Markers denote the central value of each mass bin. The width of the rectangular boxes denotes the extent of the mass bin and the height denotes the $16^{\mathrm{th}}$ and $84^{\mathrm{th}}$ percentiles on the measured multiplicity fraction by assuming it to be a binomial distribution. For clarity, only the percentiles for cases $B_0$ and $B_3$ are shown.}
\label{fig:cdf_bateratio}
\end{figure}

A final metric by which we can compare our simulations is by examining their multiplicity fractions, 
\begin{equation}
mf = \frac{B+T+Q+...}{S+B+T+Q+...},
\end{equation}
as a function of mass of the primary \citep{2005A&A...437..113H,2012MNRAS.419.3115B}. \autoref{fig:cdf_bateratio} is analogous to Figure 17 of \cite{2012MNRAS.419.3115B} and Figure 14 of \cite{2012ApJ...754...71K} where we plot the multiplicity fraction against the mass of the primary in different bound systems, including single stars. The markers denote the central values of each logarithmic mass bin and the width of the rectangular boxes for cases $B_0$ and $B_3$ denote the width of the mass bin. The height of the boxes shows the $16^{\mathrm{th}}$ and $84^{\mathrm{th}}$ percentiles of the multiplicity fraction in that bin, which we calculate using \autoref{eq:f16} and \autoref{eq:f84}. \autoref{fig:cdf_bateratio} shows that the multiplicity fraction changes as a function of the primary mass with the change in field strength. In line with what is observed in contemporary star formation \citep{2012MNRAS.419.3115B,2012ApJ...754...71K}, $mf$ increases with increasing primary mass, implying that more massive stars have more companions on average. The sharp drop in the last mass bin should be treated with caution, because it is an artifact of our choice to halt simulations at 5 per cent SFE: this guarantees by construction that all 50 $M_\odot$ stars are single. That said, we argue below that these cases likely do represent stars that will be single regardless of how far the simulation is run. Omitting these cases of very massive single stars, we find that the transition from mostly singles to mostly multiples occurs at a higher mass in the presence of a strong magnetic field; for example, we see from \autoref{fig:cdf_bateratio} that a $2\,M_{\odot}$ star is more likely to have companions in the absence of a magnetic field.

\begin{table}
\caption{KS test $p$-values for comparisons between the mass functions produced by various cases for stars of the indicated multiplicity (see main text). Thus, for example, the top row of the table means that the $p$-value we obtain by comparing the mass distribution of singles formed in case $B_0$ to those of the singles formed in cases $B_0$, $B_1$, $B_2$, and $B_3$, respectively, are 1.0 (by construction), 0.84, 0.006, and 0.02. The next row gives the corresponding values for comparing binaries in case $B_0$ to the other cases, and so forth. Low $p$-values imply that the null hypothesis that the two underlying distributions are the same can be rejected with high confidence.}
\begin{tabular}{rrrrrr}
\hline
& & $B_0$ & $B_1$ & $B_2$ & $B_3$ \\
\hline
\multirow{4}{*}{$B_0$}
& S & 1.0 & 0.84 & 0.006 & 0.02\\
& B & 1.0 & 0.12 & $10^{-5}$ & $10^{-6}$\\
& T & 1.0 & 0.32 & 0.09 & 0.22\\
& Q & 1.0 & 0.84 & 0.02 & 0.23\\
\hline
\multirow{4}{*}{$B_1$}
& S & 0.84 & 1.0 & 0.003 & 0.009 \\
& B & 0.12 & 1.0 & 0.04 & 0.02\\
& T & 0.32 & 1.0 & 0.03 & 0.23\\
& Q & 0.84 & 1.0 & 0.007 & 0.10\\
\hline
\multirow{4}{*}{$B_2$}
& S & 0.006 & 0.003 & 1.0 & 0.19\\
& B & $10^{-5}$ & 0.04 & 1.0 & 0.29\\
& T & 0.09 & 0.03 & 1.0 & 0.43\\
& Q & 0.02 & 0.007 & 1.0 & 0.66\\
\hline
\multirow{4}{*}{$B_3$}
& S & 0.02 & 0.009 & 0.19 & 1.0\\
& B & $10^{-6}$ & 0.02 & 0.29 & 1.0\\
& T & 0.22 & 0.23 & 0.43 & 1.0\\
& Q & 0.23 & 0.10 & 0.66 & 1.0\\
\hline
\end{tabular}
\label{tab:table3}
\end{table}

\section{Discussion}
\label{s:discussion}

\subsection{Implications for the primordial versus contemporary IMF}
\label{s:discuss_imf}
It is interesting to compare our results for the effects of magnetic fields on primordial star formation with work on present-day star formation, with an eye to understanding the implications for the primordial IMF. Simulations of present-day systems paint a somewhat complex picture of the effect of magnetic fields on fragmentation. While magnetic fields appear to suppress fragmentation in simulations that do not include radiative feedback \citep[e.g.,][]{2005MNRAS.362..382M,2008A&A...477...25H}, the effects in simulations that do are more subtle. Simulations of monolithic massive cores tend to find that magnetic fields also suppress fragmentation in them \citep[e.g.,][]{2010A&A...510L...3C, Myers13a}, but simulations that follow the formation of entire star clusters on larger scales generally find that magnetic fields may be less important compared to radiation feedback at suppressing fragmentation \citep[e.g.,][]{Myers14a, Cunningham18a,2019MNRAS.489.1719W}. This is because, by the time the cascade of collapse has produced a $\sim 1$ $M_\odot$ core, thermal pressure fed by radiation feedback from the protostar at the center of the core dominates over magnetic pressure \citep{2016MNRAS.460.3272K}. However, the magnetic field changes the gas distribution of the clouds already before the formation of cores starts, making the field a crucial ingredient for the initial conditions that leads to their fragmentation \citep{2008A&A...477...25H,2011ApJ...730...40P,2012MNRAS.423.2680M,2015MNRAS.450.4035F,2018PhT....71f..38F,2020IAUS..345...43F}. In this respect our finding here is more similar to the simulations of present-day massive cores, or those without radiation feedback. 

This makes sense in light of some of the important differences between present-day and population III star formation. First, the typical ``core'' that arises from cosmological simulations, and that we choose as an initial condition, is much more massive and less turbulent than a modern-day massive protostellar core, due to its higher temperature and thus lower Mach number ($\sim 1$ for a primordial core, versus $\sim 5-10$ for a colder modern-day core -- \citealt{Tan14a}). Second, due to the efficient coupling between stellar radiation and gas provided by dust grains, radiation feedback plays a much more important role in present-day star formation, and at a much earlier stage, than it does for primordial star formation. Indeed, radiation feedback appears to be the most important ingredient to suppress fragmentation in present-day star formation \citep[e.g.,][]{2009MNRAS.392.1363B, Offner09a, Krumholz11e, 2016MNRAS.460.3272K, Guszejnov16a,2017JPhCS.837a2007F}, leaving only a lesser role for magnetic fields at the later stages when cores have already formed. By contrast, in the case of primordial star formation, the inability of the gas to cool renders the entire system hotter and thus harder to affect by radiation, and the lack of solid material to couple the gas to the dust leaves a more important role for magnetic fields in primordial star formation compared to contemporary star formation.

In this light, the role of magnetic fields in shaping the primordial IMF appears to be particularly important. Even though our simulations form more than 1100 sink particles, the total number of sub-solar sinks formed in the simulations is only 15 and 6 per cent, for cases $B_0$ and $B_3$, respectively. We emphasise that this is an upper limit, since, at the time we halt the simulations, many of these low-mass sinks are still accreting, while few new fragments are emerging (see below). Thus, our simulations suggest not only that the primordial IMF was top-heavy, as has long been expected based on non-MHD arguments \citep{2006MNRAS.369..825S,2014ApJ...792...32S,2016MNRAS.462.1307S}, but that it formed a very small number of sub-solar stars that could have survived to the present day. The latter is almost entirely due to the influence of magnetic fields, which strongly suppress the disc fragmentation that otherwise seems to produce low-mass stars. In this sense magnetic fields in primordial stars appear to play the role that radiation feedback takes in present-day star formation, \textit{i.e.,} it stabilises discs and thus prevents the generation of large numbers of fragments whose masses are well below the peak of the mass function that is produced by primary fragmentation. These observations seem consistent with the fact that no surviving low-mass Population III stars have been discovered so far. 

\subsection{Caveats}
\label{s:discuss_caveats}
While our results conclusively show that magnetic fields will have an impact on the primordial IMF, there are various caveats that we should keep in mind while interpreting its implications. We list them below and comment on how significant they can be:

\begin{enumerate}
\item \textit{Resolution.}
This can be broken into two parts: the effects of resolution on fragmentation (minimum cell length that can be resolved) as well as on the number of cells per Jeans length used.
\begin{enumerate}
\item{Gravitational fragmentation.}
The sink mass distribution we analyze is resolution-dependent; at high resolution, there is a possibility that sinks with lower masses will be formed, or a different level of fragmentation will be observed. However, we showed in \cite{2019MNRAS.490..513S} that convergence (in terms of the number of sink particles) is achieved at the level of refinement we use for the same initial conditions. While the overall shape of the distribution of sink masses formed in isothermal MHD turbulence simulations is scale-free \citep{2019FrASS...6....7K}, there is no reason to believe this would be the case for non-isothermal simulations like ours where chemistry and hydrodynamics both dictate the thermodynamics of the system. However, we cannot go to higher resolutions to perform a large number of runs, due to computational limitations.
\item{Dynamo amplification.}
As we mention in \autoref{s:turb_bfields}, we use 32 cells per Jeans length as recommended by \cite{2011ApJ...731...62F} so that minimum dynamo action is resolved in the weak-field case ($B_1$), whereas we do not expect any dynamo to operate in the strong-field cases ($B_2$ and $B_3$) as they are already saturated (e.g., \citealt{2011PhRvL.107k4504F,2016JPlPh..82f5301F}). We discuss the details of the dynamo action on the core and disc in a companion paper (P. Sharda et al., in preparation); here, we simply note that, while \citet{2011ApJ...731...62F} show that 32 cells per Jeans length is the minimum resolution required to capture the dynamo at all, even at this resolution the growth rate of the dynamo is underestimated. Thus, it is possible that at higher resolution (more cells per Jeans length), case $B_1$ would behave more similarly to $B_2$ or $B_3$, since its field would grow more rapidly. However, even if this were to occur, it would in no way contradict our conclusion that primordial magnetic fields will suppress fragmentation and affect the shape of the IMF.
\end{enumerate}
\item \textit{Initial Conditions.}
While we initialise our simulations to be consistent with results from cosmological simulations and several similar simulations of primordial star formation, we cannot take into account more realistic cloud geometries, and distribution of temperature and velocity in the cloud, which can be directly derived from structure formation in the early Universe (see, for example, \citealt{2012ApJ...745..154T}). However, simulations that start from cosmological conditions are difficult to follow on scales on which the primordial clouds ultimately collapse to form the first stars. Additionally, it is highly computationally expensive to run the large number of realisations of such simulations that would be needed to perform rigorous statistical analyses to appropriately sample the IMF as we do in this work (c.f., \autoref{fig:pvalue}). Thus, there remains a trade-off between selecting more realistic initial conditions and the number of such simulations that are feasible. 
\item \textit{Radiation Feedback.}
A crucial ingredient missing in our simulations is radiation feedback. Earlier works have conclusively showed that radiation feedback can halt accretion onto massive first stars, thereby limiting their final masses \citep{2008ApJ...681..771M,2011Sci...334.1250H,2012MNRAS.422..290S,2017ApJ...835...32T,2020ApJ...892L..14S}, especially if the accretion rates are low \citep{2014ApJ...781...60H,2015MNRAS.449...77L}. This is precisely why we choose SFE = 5 per cent as our threshold beyond which we expect our results to deviate from reality. Analyses such as ours are crucial precursors to a suite of complete radiation MHD simulations of the first stars because they can disentangle the effects of magnetic fields alone.
\item \textit{Jets and Outflows.}
Jets are well known to emerge from the inner accretion disc of protostars \citep{2014prpl.conf..451F,2014prpl.conf...53O}. They can carry away mass from the protostellar-accretion disc system, thus reducing the final stellar mass and consequently impacting the IMF. \cite{2006ApJ...647L...1M} study the formation and launching of strong jets in simulations of the first stars where an ordered magnetic field is assumed, showing that the stellar mass can be reduced in such cases. However, recent present-day star formation simulations by \cite{2019MNRAS.485.5532G} show the absence of a jet in cases where the magnetic field is completely tangled and does not have an ordered component, similar to what we expect for primordial star formation at least in the early stages. Thus, jets might not have a significant impact on the primordial IMF if the field were highly tangled; however, even a very tangled field can generate an ordered component during protostellar accretion by the action of the $\alpha\omega$ dynamo \citep{1996ARA&A..34..155B,2013ApJ...778...21M,2015ApJ...808...65F}. How strong an ordered component could be generated is an open question. However, we do not have the resolution in our simulations to resolve the regions of jet launching where this might take place \citep{2014ApJ...797L..19F}.
\item \textit{Non-ideal MHD effects.}
Note that we perform ideal MHD simulations to study the role of magnetic fields during the formation of the first stars. As in the present-day Universe (e.g., \citealt{2011ApJ...736..144B,2017MNRAS.471.1488N,2018MNRAS.475.1859W,2018FrASS...5...39W}), non-ideal MHD effects are potentially important in the primordial Universe as well. For example, \cite{2009ApJ...703.1096S} and \cite{2019MNRAS.488.1846N} show that ambipolar diffusion has an impact on the thermal evolution ($n-T$) of primordial clouds. Additionally, we do not include Li as a chemical species since its importance for both chemistry and cooling have shown to be negligible \citep{2013ARA&A..51..163G,2018MNRAS.476.1826L}. However, Li has the highest ionization potential in all the primordial species and also becomes the main charge carrier at $n > 10^{8}\,\mathrm{cm^{-3}}$ \citep{2009MNRAS.393..911G}, both of which can impact the collapse. Nonetheless, \cite[see their Figure 8]{2012ApJ...754...99S} show that the growth rate of the field due to dynamo amplification is orders of magnitude more than the dissipation caused by ambipolar diffusion and Ohmic effects, except around $n \sim 10^{9}\,\mathrm{cm^{-3}}$ (see also, \citealt{2013MNRAS.435.3283M}).
\item \textit{Subsequent fragmentation and multiplicity.}
We cannot ignore the fact that we stop our simulations only a few thousand $\mathrm{yr}$ after the formation of the first sink particle. Any subsequent fragmentation that we cannot capture has the potential to change the sink mass distributions and multiplicity. The multiplicity can also change even without fragmentation as dynamical interactions take place. However, this is unlikely to alter our conclusions, because when we continue to run the simulations past the threshold, we find additional sinks appearing in only a small fraction of our runs; the number of such new sinks is less than 15 per cent of the sinks already formed in the extra time computationally feasible in which the SFE reaches 10 per cent. This is because almost all secondary fragmentation, if any, already occurs within our specified threshold, and is consistent in cases of single-star runs with the results of \cite{2013ApJ...772L...3L}.
\item \textit{Final stellar masses.}
A common and well-known feature of all simulations of Population III star formation is that they cannot be run for millions of $\mathrm{yr}$ in proper time once the stars have formed. Thus, the final masses of star particles in such simulations cannot be ascertained (see, however, \citealt{2008ApJ...681..771M,2014ApJ...781...60H}). Nonetheless, the lack of knowledge of the final stellar masses should not change the conclusions of this work; if magnetic fields already affect the distribution of Population III protostars soon after they have formed, their presence will certainly impact the distribution of their final masses.
\end{enumerate}

\section{Conclusions}
\label{s:conclusions}
In this work, we investigate how a dynamo-induced magnetic field in pre-collapse primordial clouds affects the primordial initial mass function. We do so by generating one of the largest suites of high-resolution ideal MHD simulations of the formation of the first stars. We follow 50 realizations of an isolated, initially-turbulent primordial core that only differ in the initial random turbulence and unordered magnetic field structure, for three different initial magnetic field strengths each, as motivated by various arguments for generation, sustenance and amplification of primordial magnetic fields. We also carry out control simulations where the magnetic field is absent. The 200 simulations in total form more than 1100 sink particles (used as a proxy for stars), thus providing us with a sample size sufficient to characterise the Population III IMF. 

We show that the sink mass distributions of cases with weak/zero magnetic field strength are statistically different from those produced by simulations with strong magnetic fields. We find that strong fields suppress fragmentation in primordial clouds, reducing the number of low-mass stars almost by a factor of 2. As a result of this shift, our strongly-magnetised simulations produce almost no first stars at masses $\lesssim 1$ $M_\odot$, small enough that they might be expected to survive to the present day. In contrast, in the non-magnetised cases such low-mass stars are smaller than average, but are not a very uncommon outcome. We emphasise, however, that the results of individual simulations are highly chaotic, so that sample sizes of several hundred stars are required to detect the IMF shift we observe with confidence. Studies using only a single realisation of each magnetic field strength, or even $\sim 10$ may not yield a statistically-robust signal. We also caution that, since our simulations include only an isolated core, we lack a turbulent cascade from larger scales, as would be present for fully cosmological initial conditions. We intend to explore the effects of these more realistic initial conditions in a forthcoming study.

We also find that the population of singles and binaries differ in the strong-field cases from the control simulation. The field tends to affect those simulations more where little fragmentation is present, leading to the formation of single or binary stars. In contrast, the magnetic field strength has no detectable impact on the overall clustering and multiplicity fraction of first stars. The effect we observe is simply that strong fields shift the entire mass distribution to larger values, and in the process, shift the transition where first stars go from being mostly single to mostly multiple to higher masses.

In summary, we find strong evidence that magnetic fields impact the primordial IMF to a greater extent than they do for the present-day IMF. Even with all the caveats as listed in \autoref{s:discuss_caveats}, it is clear that magnetic fields will have a significant impact on the primordial IMF, primarily by suppressing the formation of lower-mass stars even before radiation feedback kicks in to halt accretion onto massive stars. There are convincing arguments in the literature that project a strong magnetic field during the collapse of primordial clouds, due to amplification by flux-freezing and the small-scale, turbulent dynamo. Thus, future works that discuss the primordial IMF should take into account the role magnetic fields play in setting the formation and evolution of Population III stars.

\section*{Acknowledgements}
This paper was primarily written during the 2019-2020 bush fire crisis in Australia. We dedicate this work to the emergency response personnel from Australia and elsewhere who have relentlessly protected the community from severe bush fires that raged all across the country, and ensured life goes on as normal during this calamity. 

We thank the anonymous referee for their helpful suggestions. PS is supported by the Australian Government Research Training Program (RTP) Scholarship. CF and MRK acknowledge funding provided by the Australian Research Council (ARC) through Discovery Projects DP170100603 (CF) and DP190101258 (MRK) and Future Fellowships FT180100495 (CF) and FT180100375 (MRK), and the Australia-Germany Joint Research Cooperation Scheme (UA-DAAD; both CF and MRK). MRK acknowledges support from an Alexander von Humboldt Research Award. The simulations and data analyses presented in this work used high-performance computing resources provided by the Australian National Computational Infrastructure (NCI) through projects \texttt{ek9} (CF) and \texttt{jh2} (MRK) in the framework of the National Computational Merit Allocation Scheme and the Australian National University (ANU) Allocation Scheme, and as part of a contribution by NCI to the ARC Centre of Excellence for All Sky Astrophysics in 3 Dimensions (ASTRO 3D, CE170100013). Parts of this paper were also written during the ASTRO 3D writing retreat in 2019. The simulation software FLASH was in part developed by the Department of Energy$-$supported Flash Centre for Computational Science at the University of Chicago. Analysis was performed in \texttt{ipython} \citep{PER-GRA:2007} and \texttt{Jupyter} packages using \texttt{yt} \citep{2011ApJS..192....9T}, \texttt{VisIt} \citep{HPV:VisIt}, \texttt{numpy} \citep{oliphant2006guide} and \texttt{scipy} \citep{2020SciPy-NMeth}; plots were created using \texttt{Matplotlib} \citep{Hunter:2007,thomas_a_caswell_2019_3264781}. \texttt{VisIt} is supported by the Department of Energy with funding from the Advanced Simulation and Computing Program and the Scientific Discovery through Advanced Computing Program. This research has made extensive use of NASA's Astrophysics Data System Bibliographic Services. \textit{Data availability:} The simulation data underlying this article are available on request. Please email the lead author for obtaining the data.



\bibliographystyle{mnras}
\bibliography{references}

\begin{thebibliography}{}
\makeatletter
\relax
\def\mn@urlcharsother{\let\do\@makeother \do\$\do\&\do\#\do\^\do\_\do\%\do\~}
\def\mn@doi{\begingroup\mn@urlcharsother \@ifnextchar [ {\mn@doi@}
  {\mn@doi@[]}}
\def\mn@doi@[#1]#2{\def\@tempa{#1}\ifx\@tempa\@empty \href
  {http://dx.doi.org/#2} {doi:#2}\else \href {http://dx.doi.org/#2} {#1}\fi
  \endgroup}
\def\mn@eprint#1#2{\mn@eprint@#1:#2::\@nil}
\def\mn@eprint@arXiv#1{\href {http://arxiv.org/abs/#1} {{\tt arXiv:#1}}}
\def\mn@eprint@dblp#1{\href {http://dblp.uni-trier.de/rec/bibtex/#1.xml}
  {dblp:#1}}
\def\mn@eprint@#1:#2:#3:#4\@nil{\def\@tempa {#1}\def\@tempb {#2}\def\@tempc
  {#3}\ifx \@tempc \@empty \let \@tempc \@tempb \let \@tempb \@tempa \fi \ifx
  \@tempb \@empty \def\@tempb {arXiv}\fi \@ifundefined
  {mn@eprint@\@tempb}{\@tempb:\@tempc}{\expandafter \expandafter \csname
  mn@eprint@\@tempb\endcsname \expandafter{\@tempc}}}

\bibitem[\protect\citeauthoryear{{Bai} \& {Stone}}{{Bai} \&
  {Stone}}{2011}]{2011ApJ...736..144B}
{Bai} X.-N.,  {Stone} J.~M.,  2011, \mn@doi [\apj]
  {10.1088/0004-637X/736/2/144}, \href
  {https://ui.adsabs.harvard.edu/abs/2011ApJ...736..144B} {736, 144}

\bibitem[\protect\citeauthoryear{{Baierlein}}{{Baierlein}}{1978}]{1978MNRAS.184..843B}
{Baierlein} R.,  1978, \mn@doi [\mnras] {10.1093/mnras/184.4.843}, \href
  {https://ui.adsabs.harvard.edu/abs/1978MNRAS.184..843B} {184, 843}

\bibitem[\protect\citeauthoryear{{Balogh}}{{Balogh}}{2010}]{2010SSRv..152...23B}
{Balogh} A.,  2010, \mn@doi [\ssr] {10.1007/s11214-010-9643-1}, \href
  {https://ui.adsabs.harvard.edu/abs/2010SSRv..152...23B} {152, 23}

\bibitem[\protect\citeauthoryear{{Banerjee} \& {Jedamzik}}{{Banerjee} \&
  {Jedamzik}}{2004}]{2004PhRvD..70l3003B}
{Banerjee} R.,  {Jedamzik} K.,  2004, \mn@doi [\prd]
  {10.1103/PhysRevD.70.123003}, \href
  {https://ui.adsabs.harvard.edu/abs/2004PhRvD..70l3003B} {70, 123003}

\bibitem[\protect\citeauthoryear{{Bate}}{{Bate}}{2009}]{2009MNRAS.392.1363B}
{Bate} M.~R.,  2009, \mn@doi [\mnras] {10.1111/j.1365-2966.2008.14165.x}, \href
  {https://ui.adsabs.harvard.edu/abs/2009MNRAS.392.1363B} {392, 1363}

\bibitem[\protect\citeauthoryear{{Bate}}{{Bate}}{2012}]{2012MNRAS.419.3115B}
{Bate} M.~R.,  2012, \mn@doi [\mnras] {10.1111/j.1365-2966.2011.19955.x}, \href
  {https://ui.adsabs.harvard.edu/abs/2012MNRAS.419.3115B} {419, 3115}

\bibitem[\protect\citeauthoryear{{Beattie} \& {Federrath}}{{Beattie} \&
  {Federrath}}{2020}]{2020MNRAS.492..668B}
{Beattie} J.~R.,  {Federrath} C.,  2020, \mn@doi [\mnras]
  {10.1093/mnras/stz3377}, \href
  {https://ui.adsabs.harvard.edu/abs/2020MNRAS.492..668B} {492, 668}

\bibitem[\protect\citeauthoryear{{Beck}, {Brandenburg}, {Moss}, {Shukurov}  \&
  {Sokoloff}}{{Beck} et~al.}{1996}]{1996ARA&A..34..155B}
{Beck} R.,  {Brandenburg} A.,  {Moss} D.,  {Shukurov} A.,   {Sokoloff} D.,
  1996, \mn@doi [\araa] {10.1146/annurev.astro.34.1.155}, \href
  {https://ui.adsabs.harvard.edu/abs/1996ARA&A..34..155B} {34, 155}

\bibitem[\protect\citeauthoryear{{Bernet}, {Miniati}, {Lilly}, {Kronberg}  \&
  {Dessauges-Zavadsky}}{{Bernet} et~al.}{2008}]{2008Natur.454..302B}
{Bernet} M.~L.,  {Miniati} F.,  {Lilly} S.~J.,  {Kronberg} P.~P.,
  {Dessauges-Zavadsky} M.,  2008, \mn@doi [\nat] {10.1038/nature07105}, \href
  {https://ui.adsabs.harvard.edu/abs/2008Natur.454..302B} {454, 302}

\bibitem[\protect\citeauthoryear{{Bhat} \& {Subramanian}}{{Bhat} \&
  {Subramanian}}{2014}]{2014ApJ...791L..34B}
{Bhat} P.,  {Subramanian} K.,  2014, \mn@doi [\apjl]
  {10.1088/2041-8205/791/2/L34}, \href
  {https://ui.adsabs.harvard.edu/abs/2014ApJ...791L..34B} {791, L34}

\bibitem[\protect\citeauthoryear{{Biermann}}{{Biermann}}{1950}]{1950ZNatA...5...65B}
{Biermann} L.,  1950, Zeitschrift Naturforschung Teil A, \href
  {https://ui.adsabs.harvard.edu/abs/1950ZNatA...5...65B} {5, 65}

\bibitem[\protect\citeauthoryear{Bouchut, Klingenberg  \& Waagan}{Bouchut
  et~al.}{2007}]{Bouchut2007}
Bouchut F.,  Klingenberg C.,   Waagan K.,  2007, \mn@doi [Numerische
  Mathematik] {10.1007/s00211-007-0108-8}, 108, 7

\bibitem[\protect\citeauthoryear{Bouchut, Klingenberg  \& Waagan}{Bouchut
  et~al.}{2010}]{Bouchut2010}
Bouchut F.,  Klingenberg C.,   Waagan K.,  2010, \mn@doi [Numerische
  Mathematik] {10.1007/s00211-010-0289-4}, 115, 647

\bibitem[\protect\citeauthoryear{{Brandenburg} \& {Subramanian}}{{Brandenburg}
  \& {Subramanian}}{2005}]{2005PhR...417....1B}
{Brandenburg} A.,  {Subramanian} K.,  2005, \mn@doi [\physrep]
  {10.1016/j.physrep.2005.06.005}, \href
  {https://ui.adsabs.harvard.edu/abs/2005PhR...417....1B} {417, 1}

\bibitem[\protect\citeauthoryear{{Brandenburg}, {Enqvist}  \&
  {Olesen}}{{Brandenburg} et~al.}{1996}]{1996PhRvD..54.1291B}
{Brandenburg} A.,  {Enqvist} K.,   {Olesen} P.,  1996, \mn@doi [\prd]
  {10.1103/PhysRevD.54.1291}, \href
  {https://ui.adsabs.harvard.edu/abs/1996PhRvD..54.1291B} {54, 1291}

\bibitem[\protect\citeauthoryear{{Bromm}}{{Bromm}}{2013}]{2013RPPh...76k2901B}
{Bromm} V.,  2013, \mn@doi [Reports on Progress in Physics]
  {10.1088/0034-4885/76/11/112901}, \href
  {https://ui.adsabs.harvard.edu/abs/2013RPPh...76k2901B} {76, 112901}

\bibitem[\protect\citeauthoryear{{Bromm}, {Coppi}  \& {Larson}}{{Bromm}
  et~al.}{2002}]{2002ApJ...564...23B}
{Bromm} V.,  {Coppi} P.~S.,   {Larson} R.~B.,  2002, \mn@doi [\apj]
  {10.1086/323947}, \href
  {https://ui.adsabs.harvard.edu/abs/2002ApJ...564...23B} {564, 23}

\bibitem[\protect\citeauthoryear{Burgers}{Burgers}{1948}]{BURGERS1948171}
Burgers J.,  1948, Elsevier, pp 171 -- 199,
  \mn@doi{https://doi.org/10.1016/S0065-2156(08)70100-5}, \url
  {http://www.sciencedirect.com/science/article/pii/S0065215608701005}

\bibitem[\protect\citeauthoryear{{Burkert} \& {Bodenheimer}}{{Burkert} \&
  {Bodenheimer}}{1993}]{1993MNRAS.264..798B}
{Burkert} A.,  {Bodenheimer} P.,  1993, \mn@doi [\mnras]
  {10.1093/mnras/264.4.798}, \href
  {https://ui.adsabs.harvard.edu/abs/1993MNRAS.264..798B} {264, 798}

\bibitem[\protect\citeauthoryear{{Carilli} \& {Taylor}}{{Carilli} \&
  {Taylor}}{2002}]{2002ARA&A..40..319C}
{Carilli} C.~L.,  {Taylor} G.~B.,  2002, \mn@doi [\araa]
  {10.1146/annurev.astro.40.060401.093852}, \href
  {https://ui.adsabs.harvard.edu/abs/2002ARA&A..40..319C} {40, 319}

\bibitem[\protect\citeauthoryear{Caswell et~al.,}{Caswell
  et~al.}{2019}]{thomas_a_caswell_2019_3264781}
Caswell T.~A.,  et~al., 2019, matplotlib/matplotlib: REL: v3.1.1,
  \mn@doi{10.5281/zenodo.3264781}, \url
  {https://doi.org/10.5281/zenodo.3264781}

\bibitem[\protect\citeauthoryear{{Childress} \& {Gilbert}}{{Childress} \&
  {Gilbert}}{1995}]{1995stf..book.....C}
{Childress} S.,  {Gilbert} A.~D.,  1995, {Stretch, Twist, Fold}

\bibitem[\protect\citeauthoryear{Childs et~al.,}{Childs
  et~al.}{2012}]{HPV:VisIt}
Childs H.,  et~al., 2012, in , {High Performance Visualization--Enabling
  Extreme-Scale Scientific Insight}.
pp 357--372

\bibitem[\protect\citeauthoryear{{Christensson}, {Hindmarsh}  \&
  {Brandenburg}}{{Christensson} et~al.}{2001}]{2001PhRvE..64e6405C}
{Christensson} M.,  {Hindmarsh} M.,   {Brandenburg} A.,  2001, \mn@doi [\pre]
  {10.1103/PhysRevE.64.056405}, \href
  {https://ui.adsabs.harvard.edu/abs/2001PhRvE..64e6405C} {64, 056405}

\bibitem[\protect\citeauthoryear{{Clark}, {Glover}, {Klessen}  \&
  {Bromm}}{{Clark} et~al.}{2011}]{2011ApJ...727..110C}
{Clark} P.~C.,  {Glover} S. C.~O.,  {Klessen} R.~S.,   {Bromm} V.,  2011,
  \mn@doi [\apj] {10.1088/0004-637X/727/2/110}, \href
  {https://ui.adsabs.harvard.edu/abs/2011ApJ...727..110C} {727, 110}

\bibitem[\protect\citeauthoryear{{Collins}, {Kritsuk}, {Padoan}, {Li}, {Xu},
  {Ustyugov}  \& {Norman}}{{Collins} et~al.}{2012}]{2012ApJ...750...13C}
{Collins} D.~C.,  {Kritsuk} A.~G.,  {Padoan} P.,  {Li} H.,  {Xu} H.,
  {Ustyugov} S.~D.,   {Norman} M.~L.,  2012, \mn@doi [\apj]
  {10.1088/0004-637X/750/1/13}, \href
  {https://ui.adsabs.harvard.edu/abs/2012ApJ...750...13C} {750, 13}

\bibitem[\protect\citeauthoryear{{Commer{\c{c}}on}, {Hennebelle}, {Audit},
  {Chabrier}  \& {Teyssier}}{{Commer{\c{c}}on}
  et~al.}{2010}]{2010A&A...510L...3C}
{Commer{\c{c}}on} B.,  {Hennebelle} P.,  {Audit} E.,  {Chabrier} G.,
  {Teyssier} R.,  2010, \mn@doi [\aap] {10.1051/0004-6361/200913597}, \href
  {https://ui.adsabs.harvard.edu/abs/2010A&A...510L...3C} {510, L3}

\bibitem[\protect\citeauthoryear{{Crutcher}}{{Crutcher}}{2012}]{2012ARA&A..50...29C}
{Crutcher} R.~M.,  2012, \mn@doi [\araa] {10.1146/annurev-astro-081811-125514},
  \href {https://ui.adsabs.harvard.edu/abs/2012ARA&A..50...29C} {50, 29}

\bibitem[\protect\citeauthoryear{{Cunningham}, {Krumholz}, {McKee}  \&
  {Klein}}{{Cunningham} et~al.}{2018}]{Cunningham18a}
{Cunningham} A.~J.,  {Krumholz} M.~R.,  {McKee} C.~F.,   {Klein} R.~I.,  2018,
  \mn@doi [\mnras] {10.1093/mnras/sty154}, \href
  {http://adsabs.harvard.edu/abs/2018MNRAS.476..771C} {476, 771}

\bibitem[\protect\citeauthoryear{{Doi} \& {Susa}}{{Doi} \&
  {Susa}}{2011}]{2011ApJ...741...93D}
{Doi} K.,  {Susa} H.,  2011, \mn@doi [\apj] {10.1088/0004-637X/741/2/93}, \href
  {https://ui.adsabs.harvard.edu/abs/2011ApJ...741...93D} {741, 93}

\bibitem[\protect\citeauthoryear{{Donnert}, {Vazza}, {Br{\"u}ggen}  \&
  {ZuHone}}{{Donnert} et~al.}{2018}]{2018SSRv..214..122D}
{Donnert} J.,  {Vazza} F.,  {Br{\"u}ggen} M.,   {ZuHone} J.,  2018, \mn@doi
  [\ssr] {10.1007/s11214-018-0556-8}, \href
  {https://ui.adsabs.harvard.edu/abs/2018SSRv..214..122D} {214, 122}

\bibitem[\protect\citeauthoryear{{Dubey} et~al.,}{{Dubey}
  et~al.}{2008}]{2008ASPC..385..145D}
{Dubey} A.,  et~al., 2008, in {Pogorelov} N.~V.,  {Audit} E.,   {Zank} G.~P.,
  eds,  Astronomical Society of the Pacific Conference Series Vol. 385,
  Numerical Modeling of Space Plasma Flows. p.~145

\bibitem[\protect\citeauthoryear{{Dubey} et~al.,}{{Dubey}
  et~al.}{2013}]{Dubey_2013}
{Dubey} A.,  et~al., 2013, in 2013 5th International Workshop on Software
  Engineering for Computational Science and Engineering (SE-CSE). pp~1--8,
  \mn@doi{10.1109/SECSE.2013.6615093}

\bibitem[\protect\citeauthoryear{{Duffin} \& {Pudritz}}{{Duffin} \&
  {Pudritz}}{2011}]{2011IAUS..270..291D}
{Duffin} D.~F.,  {Pudritz} R.~E.,  2011, in {Alves} J.,  {Elmegreen} B.~G.,
  {Girart} J.~M.,   {Trimble} V.,  eds,  IAU Symposium Vol. 270, Computational
  Star Formation. pp 291--295 (\mn@eprint {arXiv} {1009.4445}),
  \mn@doi{10.1017/S1743921311000536}

\bibitem[\protect\citeauthoryear{{Durrer} \& {Neronov}}{{Durrer} \&
  {Neronov}}{2013}]{2013A&ARv..21...62D}
{Durrer} R.,  {Neronov} A.,  2013, \mn@doi [\aapr] {10.1007/s00159-013-0062-7},
  \href {https://ui.adsabs.harvard.edu/abs/2013A&ARv..21...62D} {21, 62}

\bibitem[\protect\citeauthoryear{{Falceta-Gon{\c{c}}alves} \&
  {Kowal}}{{Falceta-Gon{\c{c}}alves} \& {Kowal}}{2015}]{2015ApJ...808...65F}
{Falceta-Gon{\c{c}}alves} D.,  {Kowal} G.,  2015, \mn@doi [\apj]
  {10.1088/0004-637X/808/1/65}, \href
  {https://ui.adsabs.harvard.edu/abs/2015ApJ...808...65F} {808, 65}

\bibitem[\protect\citeauthoryear{{Federrath}}{{Federrath}}{2013}]{2013MNRAS.436.1245F}
{Federrath} C.,  2013, \mn@doi [\mnras] {10.1093/mnras/stt1644}, \href
  {https://ui.adsabs.harvard.edu/abs/2013MNRAS.436.1245F} {436, 1245}

\bibitem[\protect\citeauthoryear{{Federrath}}{{Federrath}}{2015}]{2015MNRAS.450.4035F}
{Federrath} C.,  2015, \mn@doi [\mnras] {10.1093/mnras/stv941}, \href
  {https://ui.adsabs.harvard.edu/abs/2015MNRAS.450.4035F} {450, 4035}

\bibitem[\protect\citeauthoryear{{Federrath}}{{Federrath}}{2016}]{2016JPlPh..82f5301F}
{Federrath} C.,  2016, \mn@doi [Journal of Plasma Physics]
  {10.1017/S0022377816001069}, \href
  {https://ui.adsabs.harvard.edu/abs/2016JPlPh..82f5301F} {82, 535820601}

\bibitem[\protect\citeauthoryear{{Federrath}}{{Federrath}}{2018}]{2018PhT....71f..38F}
{Federrath} C.,  2018, \mn@doi [Physics Today] {10.1063/PT.3.3947}, \href
  {https://ui.adsabs.harvard.edu/abs/2018PhT....71f..38F} {71, 38}

\bibitem[\protect\citeauthoryear{{Federrath}}{{Federrath}}{2020}]{2020IAUS..345...43F}
{Federrath} C.,  2020, in {Elmegreen} B.~G.,  {T{\'o}th} L.~V.,   {G{\"u}del}
  M.,  eds,  IAU Symposium Vol. 345, IAU Symposium. pp 43--50 (\mn@eprint
  {arXiv} {2002.04224}), \mn@doi{10.1017/S174392131900173X}

\bibitem[\protect\citeauthoryear{{Federrath} \& {Klessen}}{{Federrath} \&
  {Klessen}}{2012}]{2012ApJ...761..156F}
{Federrath} C.,  {Klessen} R.~S.,  2012, \mn@doi [\apj]
  {10.1088/0004-637X/761/2/156}, \href
  {https://ui.adsabs.harvard.edu/abs/2012ApJ...761..156F} {761, 156}

\bibitem[\protect\citeauthoryear{{Federrath}, {Roman-Duval}, {Klessen},
  {Schmidt}  \& {Mac Low}}{{Federrath} et~al.}{2010a}]{2010A&A...512A..81F}
{Federrath} C.,  {Roman-Duval} J.,  {Klessen} R.~S.,  {Schmidt} W.,   {Mac Low}
  M.~M.,  2010a, \mn@doi [\aap] {10.1051/0004-6361/200912437}, \href
  {https://ui.adsabs.harvard.edu/abs/2010A&A...512A..81F} {512, A81}

\bibitem[\protect\citeauthoryear{{Federrath}, {Banerjee}, {Clark}  \&
  {Klessen}}{{Federrath} et~al.}{2010b}]{2010ApJ...713..269F}
{Federrath} C.,  {Banerjee} R.,  {Clark} P.~C.,   {Klessen} R.~S.,  2010b,
  \mn@doi [\apj] {10.1088/0004-637X/713/1/269}, \href
  {https://ui.adsabs.harvard.edu/\#abs/2010ApJ...713..269F} {713, 269}

\bibitem[\protect\citeauthoryear{{Federrath}, {Chabrier}, {Schober},
  {Banerjee}, {Klessen}  \& {Schleicher}}{{Federrath}
  et~al.}{2011a}]{2011PhRvL.107k4504F}
{Federrath} C.,  {Chabrier} G.,  {Schober} J.,  {Banerjee} R.,  {Klessen}
  R.~S.,   {Schleicher} D.~R.~G.,  2011a, \mn@doi [\prl]
  {10.1103/PhysRevLett.107.114504}, \href
  {https://ui.adsabs.harvard.edu/abs/2011PhRvL.107k4504F} {107, 114504}

\bibitem[\protect\citeauthoryear{{Federrath}, {Banerjee}, {Seifried}, {Clark}
  \& {Klessen}}{{Federrath} et~al.}{2011b}]{2011IAUS..270..425F}
{Federrath} C.,  {Banerjee} R.,  {Seifried} D.,  {Clark} P.~C.,   {Klessen}
  R.~S.,  2011b, in {Alves} J.,  {Elmegreen} B.~G.,  {Girart} J.~M.,
  {Trimble} V.,  eds,  IAU Symposium Vol. 270, Computational Star Formation. pp
  425--428 (\mn@eprint {arXiv} {1007.2504}), \mn@doi{10.1017/S1743921311000755}

\bibitem[\protect\citeauthoryear{{Federrath}, {Sur}, {Schleicher}, {Banerjee}
  \& {Klessen}}{{Federrath} et~al.}{2011c}]{2011ApJ...731...62F}
{Federrath} C.,  {Sur} S.,  {Schleicher} D.~R.~G.,  {Banerjee} R.,   {Klessen}
  R.~S.,  2011c, \mn@doi [\apj] {10.1088/0004-637X/731/1/62}, \href
  {http://adsabs.harvard.edu/abs/2011ApJ...731...62F} {731, 62}

\bibitem[\protect\citeauthoryear{{Federrath}, {Schober}, {Bovino}  \&
  {Schleicher}}{{Federrath} et~al.}{2014}]{2014ApJ...797L..19F}
{Federrath} C.,  {Schober} J.,  {Bovino} S.,   {Schleicher} D. R.~G.,  2014,
  \mn@doi [\apjl] {10.1088/2041-8205/797/2/L19}, \href
  {https://ui.adsabs.harvard.edu/abs/2014ApJ...797L..19F} {797, L19}

\bibitem[\protect\citeauthoryear{{Federrath}, {Krumholz}  \&
  {Hopkins}}{{Federrath} et~al.}{2017}]{2017JPhCS.837a2007F}
{Federrath} C.,  {Krumholz} M.,   {Hopkins} P.~F.,  2017, in Journal of Physics
  Conference Series. p. 012007, \mn@doi{10.1088/1742-6596/837/1/012007}

\bibitem[\protect\citeauthoryear{{Field} \& {Carroll}}{{Field} \&
  {Carroll}}{2000}]{2000PhRvD..62j3008F}
{Field} G.~B.,  {Carroll} S.~M.,  2000, \mn@doi [\prd]
  {10.1103/PhysRevD.62.103008}, \href
  {https://ui.adsabs.harvard.edu/abs/2000PhRvD..62j3008F} {62, 103008}

\bibitem[\protect\citeauthoryear{{Forgan} \& {Rice}}{{Forgan} \&
  {Rice}}{2011}]{2011MNRAS.417.1928F}
{Forgan} D.,  {Rice} K.,  2011, \mn@doi [\mnras]
  {10.1111/j.1365-2966.2011.19380.x}, \href
  {https://ui.adsabs.harvard.edu/abs/2011MNRAS.417.1928F} {417, 1928}

\bibitem[\protect\citeauthoryear{{Frank} et~al.,}{{Frank}
  et~al.}{2014}]{2014prpl.conf..451F}
{Frank} A.,  et~al., 2014, in {Beuther} H.,  {Klessen} R.~S.,  {Dullemond}
  C.~P.,   {Henning} T.,  eds, Protostars and Planets VI. p.~451 (\mn@eprint
  {arXiv} {1402.3553}), \mn@doi{10.2458/azu_uapress_9780816531240-ch020}

\bibitem[\protect\citeauthoryear{{Fryxell} et~al.,}{{Fryxell}
  et~al.}{2000}]{2000ApJS..131..273F}
{Fryxell} B.,  et~al., 2000, \mn@doi [\apjs] {10.1086/317361}, \href
  {http://adsabs.harvard.edu/abs/2000ApJS..131..273F} {131, 273}

\bibitem[\protect\citeauthoryear{{Fumagalli}, {da Silva}  \&
  {Krumholz}}{{Fumagalli} et~al.}{2011}]{2011ApJ...741L..26F}
{Fumagalli} M.,  {da Silva} R.~L.,   {Krumholz} M.~R.,  2011, \mn@doi [\apjl]
  {10.1088/2041-8205/741/2/L26}, \href
  {https://ui.adsabs.harvard.edu/abs/2011ApJ...741L..26F} {741, L26}

\bibitem[\protect\citeauthoryear{{Galli} \& {Palla}}{{Galli} \&
  {Palla}}{2013}]{2013ARA&A..51..163G}
{Galli} D.,  {Palla} F.,  2013, \mn@doi [\araa]
  {10.1146/annurev-astro-082812-141029}, \href
  {https://ui.adsabs.harvard.edu/abs/2013ARA&A..51..163G} {51, 163}

\bibitem[\protect\citeauthoryear{{Gerola} \& {Seiden}}{{Gerola} \&
  {Seiden}}{1978}]{1978ApJ...223..129G}
{Gerola} H.,  {Seiden} P.~E.,  1978, \mn@doi [\apj] {10.1086/156243}, \href
  {https://ui.adsabs.harvard.edu/abs/1978ApJ...223..129G} {223, 129}

\bibitem[\protect\citeauthoryear{{Gerrard}, {Federrath}  \&
  {Kuruwita}}{{Gerrard} et~al.}{2019}]{2019MNRAS.485.5532G}
{Gerrard} I.~A.,  {Federrath} C.,   {Kuruwita} R.,  2019, \mn@doi [\mnras]
  {10.1093/mnras/stz784}, \href
  {https://ui.adsabs.harvard.edu/abs/2019MNRAS.485.5532G} {485, 5532}

\bibitem[\protect\citeauthoryear{{Glover} \& {Savin}}{{Glover} \&
  {Savin}}{2009}]{2009MNRAS.393..911G}
{Glover} S.~C.~O.,  {Savin} D.~W.,  2009, \mn@doi [\mnras]
  {10.1111/j.1365-2966.2008.14156.x}, \href
  {https://ui.adsabs.harvard.edu/abs/2009MNRAS.393..911G} {393, 911}

\bibitem[\protect\citeauthoryear{{Grassi}, {Bovino}, {Schleicher}  \&
  {Gianturco}}{{Grassi} et~al.}{2013}]{2013MNRAS.431.1659G}
{Grassi} T.,  {Bovino} S.,  {Schleicher} D.,   {Gianturco} F.~A.,  2013,
  \mn@doi [\mnras] {10.1093/mnras/stt284}, \href
  {http://adsabs.harvard.edu/abs/2013MNRAS.431.1659G} {431, 1659}

\bibitem[\protect\citeauthoryear{{Grassi}, {Bovino}, {Schleicher}, {Prieto},
  {Seifried}, {Simoncini}  \& {Gianturco}}{{Grassi}
  et~al.}{2014}]{2014MNRAS.439.2386G}
{Grassi} T.,  {Bovino} S.,  {Schleicher} D.~R.~G.,  {Prieto} J.,  {Seifried}
  D.,  {Simoncini} E.,   {Gianturco} F.~A.,  2014, \mn@doi [\mnras]
  {10.1093/mnras/stu114}, \href
  {http://adsabs.harvard.edu/abs/2014MNRAS.439.2386G} {439, 2386}

\bibitem[\protect\citeauthoryear{{Grasso} \& {Rubinstein}}{{Grasso} \&
  {Rubinstein}}{2001}]{2001PhR...348..163G}
{Grasso} D.,  {Rubinstein} H.~R.,  2001, \mn@doi [\physrep]
  {10.1016/S0370-1573(00)00110-1}, \href
  {https://ui.adsabs.harvard.edu/abs/2001PhR...348..163G} {348, 163}

\bibitem[\protect\citeauthoryear{{Greenstein}}{{Greenstein}}{1969}]{1969Natur.223..938G}
{Greenstein} G.,  1969, \mn@doi [\nat] {10.1038/223938b0}, \href
  {https://ui.adsabs.harvard.edu/abs/1969Natur.223..938G} {223, 938}

\bibitem[\protect\citeauthoryear{{Greif}, {Johnson}, {Klessen}  \&
  {Bromm}}{{Greif} et~al.}{2008}]{2008MNRAS.387.1021G}
{Greif} T.~H.,  {Johnson} J.~L.,  {Klessen} R.~S.,   {Bromm} V.,  2008, \mn@doi
  [\mnras] {10.1111/j.1365-2966.2008.13326.x}, \href
  {https://ui.adsabs.harvard.edu/abs/2008MNRAS.387.1021G} {387, 1021}

\bibitem[\protect\citeauthoryear{{Greif}, {White}, {Klessen}  \&
  {Springel}}{{Greif} et~al.}{2011}]{2011ApJ...736..147G}
{Greif} T.~H.,  {White} S. D.~M.,  {Klessen} R.~S.,   {Springel} V.,  2011,
  \mn@doi [\apj] {10.1088/0004-637X/736/2/147}, \href
  {https://ui.adsabs.harvard.edu/abs/2011ApJ...736..147G} {736, 147}

\bibitem[\protect\citeauthoryear{{Guszejnov}, {Krumholz}  \&
  {Hopkins}}{{Guszejnov} et~al.}{2016}]{Guszejnov16a}
{Guszejnov} D.,  {Krumholz} M.~R.,   {Hopkins} P.~F.,  2016, \mnras, \href
  {http://adsabs.harvard.edu/abs/2016MNRAS.458..673G} {458, 673}

\bibitem[\protect\citeauthoryear{{Guth}}{{Guth}}{1981}]{1981PhRvD..23..347G}
{Guth} A.~H.,  1981, \mn@doi [\prd] {10.1103/PhysRevD.23.347}, \href
  {https://ui.adsabs.harvard.edu/abs/1981PhRvD..23..347G} {23, 347}

\bibitem[\protect\citeauthoryear{{Guth} \& {Pi}}{{Guth} \&
  {Pi}}{1982}]{1982PhRvL..49.1110G}
{Guth} A.~H.,  {Pi} S.~Y.,  1982, \mn@doi [\prl] {10.1103/PhysRevLett.49.1110},
  \href {https://ui.adsabs.harvard.edu/abs/1982PhRvL..49.1110G} {49, 1110}

\bibitem[\protect\citeauthoryear{{Haemmerl{\'e}}, {Mayer}, {Klessen},
  {Hosokawa}, {Madau}  \& {Bromm}}{{Haemmerl{\'e}}
  et~al.}{2020}]{2020SSRv..216...48H}
{Haemmerl{\'e}} L.,  {Mayer} L.,  {Klessen} R.~S.,  {Hosokawa} T.,  {Madau} P.,
    {Bromm} V.,  2020, \mn@doi [\ssr] {10.1007/s11214-020-00673-y}, \href
  {https://ui.adsabs.harvard.edu/abs/2020SSRv..216...48H} {216, 48}

\bibitem[\protect\citeauthoryear{{Han}}{{Han}}{2017}]{2017ARA&A..55..111H}
{Han} J.~L.,  2017, \mn@doi [\araa] {10.1146/annurev-astro-091916-055221},
  \href {https://ui.adsabs.harvard.edu/abs/2017ARA&A..55..111H} {55, 111}

\bibitem[\protect\citeauthoryear{{Harrison}}{{Harrison}}{1970}]{1970MNRAS.147..279H}
{Harrison} E.~R.,  1970, \mn@doi [\mnras] {10.1093/mnras/147.3.279}, \href
  {https://ui.adsabs.harvard.edu/abs/1970MNRAS.147..279H} {147, 279}

\bibitem[\protect\citeauthoryear{{Haugen}, {Brandenburg}  \& {Dobler}}{{Haugen}
  et~al.}{2004}]{2004PhRvE..70a6308H}
{Haugen} N.~E.,  {Brandenburg} A.,   {Dobler} W.,  2004, \mn@doi [\pre]
  {10.1103/PhysRevE.70.016308}, \href
  {https://ui.adsabs.harvard.edu/abs/2004PhRvE..70a6308H} {70, 016308}

\bibitem[\protect\citeauthoryear{{Hennebelle} \& {Inutsuka}}{{Hennebelle} \&
  {Inutsuka}}{2019}]{2019FrASS...6....5H}
{Hennebelle} P.,  {Inutsuka} S.-i.,  2019, \mn@doi [Frontiers in Astronomy and
  Space Sciences] {10.3389/fspas.2019.00005}, \href
  {https://ui.adsabs.harvard.edu/abs/2019FrASS...6....5H} {6, 5}

\bibitem[\protect\citeauthoryear{{Hennebelle} \& {Teyssier}}{{Hennebelle} \&
  {Teyssier}}{2008}]{2008A&A...477...25H}
{Hennebelle} P.,  {Teyssier} R.,  2008, \mn@doi [\aap]
  {10.1051/0004-6361:20078310}, \href
  {https://ui.adsabs.harvard.edu/abs/2008A&A...477...25H} {477, 25}

\bibitem[\protect\citeauthoryear{{Higuchi}, {Machida}  \& {Susa}}{{Higuchi}
  et~al.}{2018}]{2018MNRAS.475.3331H}
{Higuchi} K.,  {Machida} M.~N.,   {Susa} H.,  2018, \mn@doi [\mnras]
  {10.1093/mnras/sty046}, \href
  {https://ui.adsabs.harvard.edu/abs/2018MNRAS.475.3331H} {475, 3331}

\bibitem[\protect\citeauthoryear{{Hirano}, {Hosokawa}, {Yoshida}, {Umeda},
  {Omukai}, {Chiaki}  \& {Yorke}}{{Hirano} et~al.}{2014}]{2014ApJ...781...60H}
{Hirano} S.,  {Hosokawa} T.,  {Yoshida} N.,  {Umeda} H.,  {Omukai} K.,
  {Chiaki} G.,   {Yorke} H.~W.,  2014, \mn@doi [\apj]
  {10.1088/0004-637X/781/2/60}, \href
  {https://ui.adsabs.harvard.edu/abs/2014ApJ...781...60H} {781, 60}

\bibitem[\protect\citeauthoryear{{Hirano}, {Hosokawa}, {Yoshida}, {Omukai}  \&
  {Yorke}}{{Hirano} et~al.}{2015}]{2015MNRAS.448..568H}
{Hirano} S.,  {Hosokawa} T.,  {Yoshida} N.,  {Omukai} K.,   {Yorke} H.~W.,
  2015, \mn@doi [\mnras] {10.1093/mnras/stv044}, \href
  {https://ui.adsabs.harvard.edu/abs/2015MNRAS.448..568H} {448, 568}

\bibitem[\protect\citeauthoryear{{Hopkins}}{{Hopkins}}{2013}]{2013MNRAS.430.1653H}
{Hopkins} P.~F.,  2013, \mn@doi [\mnras] {10.1093/mnras/sts704}, \href
  {https://ui.adsabs.harvard.edu/abs/2013MNRAS.430.1653H} {430, 1653}

\bibitem[\protect\citeauthoryear{{Hopkins} \& {Christiansen}}{{Hopkins} \&
  {Christiansen}}{2013}]{2013ApJ...776...48H}
{Hopkins} P.~F.,  {Christiansen} J.~L.,  2013, \mn@doi [\apj]
  {10.1088/0004-637X/776/1/48}, \href
  {https://ui.adsabs.harvard.edu/abs/2013ApJ...776...48H} {776, 48}

\bibitem[\protect\citeauthoryear{{Hosking} \& {Whitworth}}{{Hosking} \&
  {Whitworth}}{2004}]{2004MNRAS.347.1001H}
{Hosking} J.~G.,  {Whitworth} A.~P.,  2004, \mn@doi [\mnras]
  {10.1111/j.1365-2966.2004.07274.x}, \href
  {https://ui.adsabs.harvard.edu/abs/2004MNRAS.347.1001H} {347, 1001}

\bibitem[\protect\citeauthoryear{{Hosokawa}, {Omukai}, {Yoshida}  \&
  {Yorke}}{{Hosokawa} et~al.}{2011}]{2011Sci...334.1250H}
{Hosokawa} T.,  {Omukai} K.,  {Yoshida} N.,   {Yorke} H.~W.,  2011, \mn@doi
  [Science] {10.1126/science.1207433}, \href
  {https://ui.adsabs.harvard.edu/abs/2011Sci...334.1250H} {334, 1250}

\bibitem[\protect\citeauthoryear{{Hubber} \& {Whitworth}}{{Hubber} \&
  {Whitworth}}{2005}]{2005A&A...437..113H}
{Hubber} D.~A.,  {Whitworth} A.~P.,  2005, \mn@doi [\aap]
  {10.1051/0004-6361:20042428}, \href
  {https://ui.adsabs.harvard.edu/abs/2005A&A...437..113H} {437, 113}

\bibitem[\protect\citeauthoryear{Hunter}{Hunter}{2007}]{Hunter:2007}
Hunter J.~D.,  2007, \mn@doi [Computing in Science \& Engineering]
  {10.1109/MCSE.2007.55}, 9, 90

\bibitem[\protect\citeauthoryear{{Hutschenreuter}, {Dorn}, {Jasche}, {Vazza},
  {Paoletti}, {Lavaux}  \& {En{\ss}lin}}{{Hutschenreuter}
  et~al.}{2018}]{2018CQGra..35o4001H}
{Hutschenreuter} S.,  {Dorn} S.,  {Jasche} J.,  {Vazza} F.,  {Paoletti} D.,
  {Lavaux} G.,   {En{\ss}lin} T.~A.,  2018, \mn@doi [Classical and Quantum
  Gravity] {10.1088/1361-6382/aacde0}, \href
  {https://ui.adsabs.harvard.edu/abs/2018CQGra..35o4001H} {35, 154001}

\bibitem[\protect\citeauthoryear{{Iapichino}, {Adamek}, {Schmidt}  \&
  {Niemeyer}}{{Iapichino} et~al.}{2008}]{2008MNRAS.388.1079I}
{Iapichino} L.,  {Adamek} J.,  {Schmidt} W.,   {Niemeyer} J.~C.,  2008, \mn@doi
  [\mnras] {10.1111/j.1365-2966.2008.13137.x}, \href
  {https://ui.adsabs.harvard.edu/abs/2008MNRAS.388.1079I} {388, 1079}

\bibitem[\protect\citeauthoryear{{Iapichino}, {Federrath}  \&
  {Klessen}}{{Iapichino} et~al.}{2017}]{2017MNRAS.469.3641I}
{Iapichino} L.,  {Federrath} C.,   {Klessen} R.~S.,  2017, \mn@doi [\mnras]
  {10.1093/mnras/stx882}, \href
  {https://ui.adsabs.harvard.edu/abs/2017MNRAS.469.3641I} {469, 3641}

\bibitem[\protect\citeauthoryear{{Ichiki}, {Takahashi}, {Ohno}, {Hanayama}  \&
  {Sugiyama}}{{Ichiki} et~al.}{2006}]{2006Sci...311..827I}
{Ichiki} K.,  {Takahashi} K.,  {Ohno} H.,  {Hanayama} H.,   {Sugiyama} N.,
  2006, \mn@doi [Science] {10.1126/science.1120690}, \href
  {https://ui.adsabs.harvard.edu/abs/2006Sci...311..827I} {311, 827}

\bibitem[\protect\citeauthoryear{{Jones}, {Porter}, {Ryu}  \& {Cho}}{{Jones}
  et~al.}{2011}]{2011MmSAI..82..588J}
{Jones} T.~W.,  {Porter} D.~H.,  {Ryu} D.,   {Cho} J.,  2011, \memsai, \href
  {https://ui.adsabs.harvard.edu/abs/2011MmSAI..82..588J} {82, 588}

\bibitem[\protect\citeauthoryear{{Kahniashvili}, {Tevzadze}, {Brand enburg}  \&
  {Neronov}}{{Kahniashvili} et~al.}{2013a}]{2013PhRvD..87h3007K}
{Kahniashvili} T.,  {Tevzadze} A.~G.,  {Brand enburg} A.,   {Neronov} A.,
  2013a, \mn@doi [\prd] {10.1103/PhysRevD.87.083007}, \href
  {https://ui.adsabs.harvard.edu/abs/2013PhRvD..87h3007K} {87, 083007}

\bibitem[\protect\citeauthoryear{{Kahniashvili}, {Maravin}, {Natarajan},
  {Battaglia}  \& {Tevzadze}}{{Kahniashvili}
  et~al.}{2013b}]{2013ApJ...770...47K}
{Kahniashvili} T.,  {Maravin} Y.,  {Natarajan} A.,  {Battaglia} N.,
  {Tevzadze} A.~G.,  2013b, \mn@doi [\apj] {10.1088/0004-637X/770/1/47}, \href
  {https://ui.adsabs.harvard.edu/abs/2013ApJ...770...47K} {770, 47}

\bibitem[\protect\citeauthoryear{{Kauffmann}, {Heckman}, {De Lucia},
  {Brinchmann}, {Charlot}, {Tremonti}, {White}  \& {Brinkmann}}{{Kauffmann}
  et~al.}{2006}]{2006MNRAS.367.1394K}
{Kauffmann} G.,  {Heckman} T.~M.,  {De Lucia} G.,  {Brinchmann} J.,  {Charlot}
  S.,  {Tremonti} C.,  {White} S. D.~M.,   {Brinkmann} J.,  2006, \mn@doi
  [\mnras] {10.1111/j.1365-2966.2006.10061.x}, \href
  {https://ui.adsabs.harvard.edu/abs/2006MNRAS.367.1394K} {367, 1394}

\bibitem[\protect\citeauthoryear{{Kawasaki} \& {Kusakabe}}{{Kawasaki} \&
  {Kusakabe}}{2012}]{2012PhRvD..86f3003K}
{Kawasaki} M.,  {Kusakabe} M.,  2012, \mn@doi [\prd]
  {10.1103/PhysRevD.86.063003}, \href
  {https://ui.adsabs.harvard.edu/abs/2012PhRvD..86f3003K} {86, 063003}

\bibitem[\protect\citeauthoryear{{Kazantsev}}{{Kazantsev}}{1968}]{1968JETP...26.1031K}
{Kazantsev} A.~P.,  1968, Soviet Journal of Experimental and Theoretical
  Physics, \href {https://ui.adsabs.harvard.edu/abs/1968JETP...26.1031K} {26,
  1031}

\bibitem[\protect\citeauthoryear{{Kazantsev}, {Ruzmaikin}  \&
  {Sokolov}}{{Kazantsev} et~al.}{1985}]{1985ZhETF..88..487K}
{Kazantsev} A.~P.,  {Ruzmaikin} A.~A.,   {Sokolov} D.~D.,  1985, Zhurnal
  Eksperimentalnoi i Teoreticheskoi Fiziki, \href
  {https://ui.adsabs.harvard.edu/abs/1985ZhETF..88..487K} {88, 487}

\bibitem[\protect\citeauthoryear{{Kibble}}{{Kibble}}{1980}]{1980PhR....67..183K}
{Kibble} T.~W.~B.,  1980, \mn@doi [\physrep] {10.1016/0370-1573(80)90091-5},
  \href {https://ui.adsabs.harvard.edu/abs/1980PhR....67..183K} {67, 183}

\bibitem[\protect\citeauthoryear{{Kim}, {Olinto}  \& {Rosner}}{{Kim}
  et~al.}{1996}]{1996ApJ...468...28K}
{Kim} E.-J.,  {Olinto} A.~V.,   {Rosner} R.,  1996, \mn@doi [\apj]
  {10.1086/177667}, \href
  {https://ui.adsabs.harvard.edu/abs/1996ApJ...468...28K} {468, 28}

\bibitem[\protect\citeauthoryear{{Klessen}}{{Klessen}}{2019}]{2019ffbh.book...67K}
{Klessen} R.,  2019, {Formation of the first stars}.
pp 67--97, \mn@doi{10.1142/9789813227958_0004}

\bibitem[\protect\citeauthoryear{{Kolmogorov}}{{Kolmogorov}}{1941}]{1941DoSSR..32...16K}
{Kolmogorov} A.~N.,  1941, Akademiia Nauk SSSR Doklady, \href
  {https://ui.adsabs.harvard.edu/abs/1941DoSSR..32...16K} {32, 16}

\bibitem[\protect\citeauthoryear{{Kritsuk}, {Norman}  \& {Padoan}}{{Kritsuk}
  et~al.}{2006}]{2006ApJ...638L..25K}
{Kritsuk} A.~G.,  {Norman} M.~L.,   {Padoan} P.,  2006, \mn@doi [\apjl]
  {10.1086/500688}, \href
  {https://ui.adsabs.harvard.edu/abs/2006ApJ...638L..25K} {638, L25}

\bibitem[\protect\citeauthoryear{{Kritsuk}, {Norman}, {Padoan}  \&
  {Wagner}}{{Kritsuk} et~al.}{2007}]{2007ApJ...665..416K}
{Kritsuk} A.~G.,  {Norman} M.~L.,  {Padoan} P.,   {Wagner} R.,  2007, \mn@doi
  [\apj] {10.1086/519443}, \href
  {https://ui.adsabs.harvard.edu/abs/2007ApJ...665..416K} {665, 416}

\bibitem[\protect\citeauthoryear{{Kritsuk}, {Ustyugov}, {Norman}  \&
  {Padoan}}{{Kritsuk} et~al.}{2009}]{2009JPhCS.180a2020K}
{Kritsuk} A.~G.,  {Ustyugov} S.~D.,  {Norman} M.~L.,   {Padoan} P.,  2009, in
  Journal of Physics Conference Series. p. 012020 (\mn@eprint {arXiv}
  {0908.0378}), \mn@doi{10.1088/1742-6596/180/1/012020}

\bibitem[\protect\citeauthoryear{{Kritsuk} et~al.,}{{Kritsuk}
  et~al.}{2011}]{2011ApJ...737...13K}
{Kritsuk} A.~G.,  et~al., 2011, \mn@doi [\apj] {10.1088/0004-637X/737/1/13},
  \href {https://ui.adsabs.harvard.edu/abs/2011ApJ...737...13K} {737, 13}

\bibitem[\protect\citeauthoryear{{Kronberg}, {Bernet}, {Miniati}, {Lilly},
  {Short}  \& {Higdon}}{{Kronberg} et~al.}{2008}]{2008ApJ...676...70K}
{Kronberg} P.~P.,  {Bernet} M.~L.,  {Miniati} F.,  {Lilly} S.~J.,  {Short}
  M.~B.,   {Higdon} D.~M.,  2008, \mn@doi [\apj] {10.1086/527281}, \href
  {https://ui.adsabs.harvard.edu/abs/2008ApJ...676...70K} {676, 70}

\bibitem[\protect\citeauthoryear{{Krumholz}}{{Krumholz}}{2011}]{Krumholz11e}
{Krumholz} M.~R.,  2011, \mn@doi [\apj] {10.1088/0004-637X/743/2/110}, \href
  {http://adsabs.harvard.edu/abs/2011ApJ...743..110K} {743, 110}

\bibitem[\protect\citeauthoryear{{Krumholz}}{{Krumholz}}{2014}]{2014PhR...539...49K}
{Krumholz} M.~R.,  2014, \mn@doi [\physrep] {10.1016/j.physrep.2014.02.001},
  \href {https://ui.adsabs.harvard.edu/abs/2014PhR...539...49K} {539, 49}

\bibitem[\protect\citeauthoryear{{Krumholz} \& {Federrath}}{{Krumholz} \&
  {Federrath}}{2019}]{2019FrASS...6....7K}
{Krumholz} M.~R.,  {Federrath} C.,  2019, \mn@doi [Frontiers in Astronomy and
  Space Sciences] {10.3389/fspas.2019.00007}, \href
  {https://ui.adsabs.harvard.edu/abs/2019FrASS...6....7K} {6, 7}

\bibitem[\protect\citeauthoryear{{Krumholz}, {Klein}  \& {McKee}}{{Krumholz}
  et~al.}{2012}]{2012ApJ...754...71K}
{Krumholz} M.~R.,  {Klein} R.~I.,   {McKee} C.~F.,  2012, \mn@doi [\apj]
  {10.1088/0004-637X/754/1/71}, \href
  {https://ui.adsabs.harvard.edu/abs/2012ApJ...754...71K} {754, 71}

\bibitem[\protect\citeauthoryear{{Krumholz}, {Myers}, {Klein}  \&
  {McKee}}{{Krumholz} et~al.}{2016}]{2016MNRAS.460.3272K}
{Krumholz} M.~R.,  {Myers} A.~T.,  {Klein} R.~I.,   {McKee} C.~F.,  2016,
  \mn@doi [\mnras] {10.1093/mnras/stw1236}, \href
  {https://ui.adsabs.harvard.edu/abs/2016MNRAS.460.3272K} {460, 3272}

\bibitem[\protect\citeauthoryear{{Kulsrud}}{{Kulsrud}}{1999}]{1999ARA&A..37...37K}
{Kulsrud} R.~M.,  1999, \mn@doi [\araa] {10.1146/annurev.astro.37.1.37}, \href
  {https://ui.adsabs.harvard.edu/abs/1999ARA&A..37...37K} {37, 37}

\bibitem[\protect\citeauthoryear{{Kuruwita} \& {Federrath}}{{Kuruwita} \&
  {Federrath}}{2019}]{2019MNRAS.486.3647K}
{Kuruwita} R.~L.,  {Federrath} C.,  2019, \mn@doi [\mnras]
  {10.1093/mnras/stz1053}, \href
  {https://ui.adsabs.harvard.edu/abs/2019MNRAS.486.3647K} {486, 3647}

\bibitem[\protect\citeauthoryear{{Latif} \& {Schleicher}}{{Latif} \&
  {Schleicher}}{2015}]{2015MNRAS.449...77L}
{Latif} M.~A.,  {Schleicher} D.~R.~G.,  2015, \mn@doi [\mnras]
  {10.1093/mnras/stu2573}, \href
  {https://ui.adsabs.harvard.edu/abs/2015MNRAS.449...77L} {449, 77}

\bibitem[\protect\citeauthoryear{{Latif} \& {Schleicher}}{{Latif} \&
  {Schleicher}}{2016}]{2016A&A...585A.151L}
{Latif} M.~A.,  {Schleicher} D.~R.~G.,  2016, \mn@doi [\aap]
  {10.1051/0004-6361/201527266}, \href
  {https://ui.adsabs.harvard.edu/abs/2016A&A...585A.151L} {585, A151}

\bibitem[\protect\citeauthoryear{{Latif}, {Schleicher}, {Schmidt}  \&
  {Niemeyer}}{{Latif} et~al.}{2013a}]{2013MNRAS.432..668L}
{Latif} M.~A.,  {Schleicher} D.~R.~G.,  {Schmidt} W.,   {Niemeyer} J.,  2013a,
  \mn@doi [\mnras] {10.1093/mnras/stt503}, \href
  {https://ui.adsabs.harvard.edu/abs/2013MNRAS.432..668L} {432, 668}

\bibitem[\protect\citeauthoryear{{Latif}, {Schleicher}, {Schmidt}  \&
  {Niemeyer}}{{Latif} et~al.}{2013b}]{2013ApJ...772L...3L}
{Latif} M.~A.,  {Schleicher} D.~R.~G.,  {Schmidt} W.,   {Niemeyer} J.,  2013b,
  \mn@doi [\apjl] {10.1088/2041-8205/772/1/L3}, \href
  {https://ui.adsabs.harvard.edu/abs/2013ApJ...772L...3L} {772, L3}

\bibitem[\protect\citeauthoryear{{Latif}, {Schleicher}  \& {Schmidt}}{{Latif}
  et~al.}{2014}]{2014MNRAS.440.1551L}
{Latif} M.~A.,  {Schleicher} D.~R.~G.,   {Schmidt} W.,  2014, \mn@doi [\mnras]
  {10.1093/mnras/stu357}, \href
  {https://ui.adsabs.harvard.edu/abs/2014MNRAS.440.1551L} {440, 1551}

\bibitem[\protect\citeauthoryear{{Lemaster} \& {Stone}}{{Lemaster} \&
  {Stone}}{2009}]{2009ApJ...691.1092L}
{Lemaster} M.~N.,  {Stone} J.~M.,  2009, \mn@doi [\apj]
  {10.1088/0004-637X/691/2/1092}, \href
  {https://ui.adsabs.harvard.edu/abs/2009ApJ...691.1092L} {691, 1092}

\bibitem[\protect\citeauthoryear{{Liao}, {Turk}  \& {Schive}}{{Liao}
  et~al.}{2019}]{2019arXiv191107898L}
{Liao} W.-T.,  {Turk} M.,   {Schive} H.-Y.,  2019, arXiv e-prints, \href
  {https://ui.adsabs.harvard.edu/abs/2019arXiv191107898L} {p. arXiv:1911.07898}

\bibitem[\protect\citeauthoryear{{Liu} \& {Bromm}}{{Liu} \&
  {Bromm}}{2018}]{2018MNRAS.476.1826L}
{Liu} B.,  {Bromm} V.,  2018, \mn@doi [\mnras] {10.1093/mnras/sty350}, \href
  {https://ui.adsabs.harvard.edu/abs/2018MNRAS.476.1826L} {476, 1826}

\bibitem[\protect\citeauthoryear{{Machida} \& {Doi}}{{Machida} \&
  {Doi}}{2013}]{2013MNRAS.435.3283M}
{Machida} M.~N.,  {Doi} K.,  2013, \mn@doi [\mnras] {10.1093/mnras/stt1524},
  \href {https://ui.adsabs.harvard.edu/abs/2013MNRAS.435.3283M} {435, 3283}

\bibitem[\protect\citeauthoryear{{Machida}, {Matsumoto}, {Hanawa}  \&
  {Tomisaka}}{{Machida} et~al.}{2005}]{2005MNRAS.362..382M}
{Machida} M.~N.,  {Matsumoto} T.,  {Hanawa} T.,   {Tomisaka} K.,  2005, \mn@doi
  [\mnras] {10.1111/j.1365-2966.2005.09327.x}, \href
  {https://ui.adsabs.harvard.edu/abs/2005MNRAS.362..382M} {362, 382}

\bibitem[\protect\citeauthoryear{{Machida}, {Omukai}, {Matsumoto}  \&
  {Inutsuka}}{{Machida} et~al.}{2006}]{2006ApJ...647L...1M}
{Machida} M.~N.,  {Omukai} K.,  {Matsumoto} T.,   {Inutsuka} S.-i.,  2006,
  \mn@doi [\apjl] {10.1086/507326}, \href
  {https://ui.adsabs.harvard.edu/abs/2006ApJ...647L...1M} {647, L1}

\bibitem[\protect\citeauthoryear{{Machida}, {Omukai}, {Matsumoto}  \&
  {Inutsuka}}{{Machida} et~al.}{2008a}]{2008ApJ...677..813M}
{Machida} M.~N.,  {Omukai} K.,  {Matsumoto} T.,   {Inutsuka} S.-i.,  2008a,
  \mn@doi [\apj] {10.1086/533434}, \href
  {https://ui.adsabs.harvard.edu/abs/2008ApJ...677..813M} {677, 813}

\bibitem[\protect\citeauthoryear{{Machida}, {Matsumoto}  \&
  {Inutsuka}}{{Machida} et~al.}{2008b}]{2008ApJ...685..690M}
{Machida} M.~N.,  {Matsumoto} T.,   {Inutsuka} S.-i.,  2008b, \mn@doi [\apj]
  {10.1086/591074}, \href
  {https://ui.adsabs.harvard.edu/abs/2008ApJ...685..690M} {685, 690}

\bibitem[\protect\citeauthoryear{{Maio}, {Koopmans}  \& {Ciardi}}{{Maio}
  et~al.}{2011}]{2011MNRAS.412L..40M}
{Maio} U.,  {Koopmans} L. V.~E.,   {Ciardi} B.,  2011, \mn@doi [\mnras]
  {10.1111/j.1745-3933.2010.01001.x}, \href
  {https://ui.adsabs.harvard.edu/abs/2011MNRAS.412L..40M} {412, L40}

\bibitem[\protect\citeauthoryear{{Maki} \& {Susa}}{{Maki} \&
  {Susa}}{2004}]{2004ApJ...609..467M}
{Maki} H.,  {Susa} H.,  2004, \mn@doi [\apj] {10.1086/421103}, \href
  {https://ui.adsabs.harvard.edu/abs/2004ApJ...609..467M} {609, 467}

\bibitem[\protect\citeauthoryear{{Maki} \& {Susa}}{{Maki} \&
  {Susa}}{2007}]{2007PASJ...59..787M}
{Maki} H.,  {Susa} H.,  2007, \mn@doi [\pasj] {10.1093/pasj/59.4.787}, \href
  {https://ui.adsabs.harvard.edu/abs/2007PASJ...59..787M} {59, 787}

\bibitem[\protect\citeauthoryear{{Malapaka} \& {M{\"u}ller}}{{Malapaka} \&
  {M{\"u}ller}}{2013}]{2013ApJ...778...21M}
{Malapaka} S.~K.,  {M{\"u}ller} W.-C.,  2013, \mn@doi [\apj]
  {10.1088/0004-637X/778/1/21}, \href
  {https://ui.adsabs.harvard.edu/abs/2013ApJ...778...21M} {778, 21}

\bibitem[\protect\citeauthoryear{{Matese} \& {O'Connell}}{{Matese} \&
  {O'Connell}}{1970}]{1970ApJ...160..451M}
{Matese} J.~J.,  {O'Connell} R.~F.,  1970, \mn@doi [\apj] {10.1086/150446},
  \href {https://ui.adsabs.harvard.edu/abs/1970ApJ...160..451M} {160, 451}

\bibitem[\protect\citeauthoryear{{McKee} \& {Ostriker}}{{McKee} \&
  {Ostriker}}{2007}]{2007ARA&A..45..565M}
{McKee} C.~F.,  {Ostriker} E.~C.,  2007, \mn@doi [\araa]
  {10.1146/annurev.astro.45.051806.110602}, \href
  {https://ui.adsabs.harvard.edu/abs/2007ARA&A..45..565M} {45, 565}

\bibitem[\protect\citeauthoryear{{McKee} \& {Tan}}{{McKee} \&
  {Tan}}{2008}]{2008ApJ...681..771M}
{McKee} C.~F.,  {Tan} J.~C.,  2008, \mn@doi [\apj] {10.1086/587434}, \href
  {https://ui.adsabs.harvard.edu/abs/2008ApJ...681..771M} {681, 771}

\bibitem[\protect\citeauthoryear{{Mielke}, {Peterson}, {Schwenke}, {Garrett},
  {Truhlar}, {Michael}, {Su}  \& {Sutherland}}{{Mielke}
  et~al.}{2003}]{2003PhRvL..91f3201M}
{Mielke} S.~L.,  {Peterson} K.~A.,  {Schwenke} D.~W.,  {Garrett} B.~C.,
  {Truhlar} D.~G.,  {Michael} J.~V.,  {Su} M.-C.,   {Sutherland} J.~W.,  2003,
  \mn@doi [\prl] {10.1103/PhysRevLett.91.063201}, \href
  {https://ui.adsabs.harvard.edu/abs/2003PhRvL..91f3201M} {91, 063201}

\bibitem[\protect\citeauthoryear{{Molina}, {Glover}, {Federrath}  \&
  {Klessen}}{{Molina} et~al.}{2012}]{2012MNRAS.423.2680M}
{Molina} F.~Z.,  {Glover} S.~C.~O.,  {Federrath} C.,   {Klessen} R.~S.,  2012,
  \mn@doi [\mnras] {10.1111/j.1365-2966.2012.21075.x}, \href
  {https://ui.adsabs.harvard.edu/abs/2012MNRAS.423.2680M} {423, 2680}

\bibitem[\protect\citeauthoryear{{Mosquera Cuesta} \& {Lambiase}}{{Mosquera
  Cuesta} \& {Lambiase}}{2009}]{2009PhRvD..80b3013M}
{Mosquera Cuesta} H.~J.,  {Lambiase} G.,  2009, \mn@doi [\prd]
  {10.1103/PhysRevD.80.023013}, \href
  {https://ui.adsabs.harvard.edu/abs/2009PhRvD..80b3013M} {80, 023013}

\bibitem[\protect\citeauthoryear{{Myers}, {McKee}, {Cunningham}, {Klein}  \&
  {Krumholz}}{{Myers} et~al.}{2013}]{Myers13a}
{Myers} A.~T.,  {McKee} C.~F.,  {Cunningham} A.~J.,  {Klein} R.~I.,
  {Krumholz} M.~R.,  2013, \mn@doi [\apj] {10.1088/0004-637X/766/2/97}, \href
  {http://adsabs.harvard.edu/abs/2013ApJ...766...97M} {766, 97}

\bibitem[\protect\citeauthoryear{{Myers}, {Klein}, {Krumholz}  \&
  {McKee}}{{Myers} et~al.}{2014}]{Myers14a}
{Myers} A.~T.,  {Klein} R.~I.,  {Krumholz} M.~R.,   {McKee} C.~F.,  2014,
  \mn@doi [\mnras] {10.1093/mnras/stu190}, \href
  {http://adsabs.harvard.edu/abs/2014MNRAS.439.3420M} {439, 3420}

\bibitem[\protect\citeauthoryear{{Nakamura} \& {Umemura}}{{Nakamura} \&
  {Umemura}}{2002}]{2002ApJ...569..549N}
{Nakamura} F.,  {Umemura} M.,  2002, \mn@doi [\apj] {10.1086/339392}, \href
  {https://ui.adsabs.harvard.edu/abs/2002ApJ...569..549N} {569, 549}

\bibitem[\protect\citeauthoryear{{Nakauchi}, {Omukai}  \& {Susa}}{{Nakauchi}
  et~al.}{2019}]{2019MNRAS.488.1846N}
{Nakauchi} D.,  {Omukai} K.,   {Susa} H.,  2019, \mn@doi [\mnras]
  {10.1093/mnras/stz1799}, \href
  {https://ui.adsabs.harvard.edu/abs/2019MNRAS.488.1846N} {488, 1846}

\bibitem[\protect\citeauthoryear{{Ng} \& {Vachaspati}}{{Ng} \&
  {Vachaspati}}{2010}]{2010PhRvD..82b3008N}
{Ng} Y.,  {Vachaspati} T.,  2010, \mn@doi [\prd] {10.1103/PhysRevD.82.023008},
  \href {https://ui.adsabs.harvard.edu/abs/2010PhRvD..82b3008N} {82, 023008}

\bibitem[\protect\citeauthoryear{{Nolan}, {Salmeron}, {Federrath}, {Bicknell}
  \& {Sutherland}}{{Nolan} et~al.}{2017}]{2017MNRAS.471.1488N}
{Nolan} C.~A.,  {Salmeron} R.,  {Federrath} C.,  {Bicknell} G.~V.,
  {Sutherland} R.~S.,  2017, \mn@doi [\mnras] {10.1093/mnras/stx1642}, \href
  {https://ui.adsabs.harvard.edu/abs/2017MNRAS.471.1488N} {471, 1488}

\bibitem[\protect\citeauthoryear{{Offner}, {Klein}, {McKee}  \&
  {Krumholz}}{{Offner} et~al.}{2009}]{Offner09a}
{Offner} S.~S.~R.,  {Klein} R.~I.,  {McKee} C.~F.,   {Krumholz} M.~R.,  2009,
  \mn@doi [\apj] {10.1088/0004-637X/703/1/131}, \href
  {http://adsabs.harvard.edu/abs/2009ApJ...703..131O} {703, 131}

\bibitem[\protect\citeauthoryear{{Offner}, {Clark}, {Hennebelle}, {Bastian},
  {Bate}, {Hopkins}, {Moraux}  \& {Whitworth}}{{Offner}
  et~al.}{2014}]{2014prpl.conf...53O}
{Offner} S.~S.~R.,  {Clark} P.~C.,  {Hennebelle} P.,  {Bastian} N.,  {Bate}
  M.~R.,  {Hopkins} P.~F.,  {Moraux} E.,   {Whitworth} A.~P.,  2014, in
  {Beuther} H.,  {Klessen} R.~S.,  {Dullemond} C.~P.,   {Henning} T.,  eds,
  Protostars and Planets VI. p.~53 (\mn@eprint {arXiv} {1312.5326}),
  \mn@doi{10.2458/azu_uapress_9780816531240-ch003}

\bibitem[\protect\citeauthoryear{Oliphant}{Oliphant}{2006}]{oliphant2006guide}
Oliphant T.~E.,  2006, A guide to NumPy.
 Vol. 1, Trelgol Publishing USA

\bibitem[\protect\citeauthoryear{{Omukai} \& {Nishi}}{{Omukai} \&
  {Nishi}}{1998}]{1998ApJ...508..141O}
{Omukai} K.,  {Nishi} R.,  1998, \mn@doi [\apj] {10.1086/306395}, \href
  {https://ui.adsabs.harvard.edu/abs/1998ApJ...508..141O} {508, 141}

\bibitem[\protect\citeauthoryear{{Omukai}, {Tsuribe}, {Schneider}  \&
  {Ferrara}}{{Omukai} et~al.}{2005}]{2005ApJ...626..627O}
{Omukai} K.,  {Tsuribe} T.,  {Schneider} R.,   {Ferrara} A.,  2005, \mn@doi
  [\apj] {10.1086/429955}, \href
  {https://ui.adsabs.harvard.edu/abs/2005ApJ...626..627O} {626, 627}

\bibitem[\protect\citeauthoryear{{Padoan} \& {Nordlund}}{{Padoan} \&
  {Nordlund}}{2011}]{2011ApJ...730...40P}
{Padoan} P.,  {Nordlund} {\r{A}}.,  2011, \mn@doi [\apj]
  {10.1088/0004-637X/730/1/40}, \href
  {https://ui.adsabs.harvard.edu/abs/2011ApJ...730...40P} {730, 40}

\bibitem[\protect\citeauthoryear{P\'erez \& Granger}{P\'erez \&
  Granger}{2007}]{PER-GRA:2007}
P\'erez F.,  Granger B.~E.,  2007, \mn@doi [Computing in Science and
  Engineering] {10.1109/MCSE.2007.53}, 9, 21

\bibitem[\protect\citeauthoryear{{Price} \& {Bate}}{{Price} \&
  {Bate}}{2007}]{2007MNRAS.377...77P}
{Price} D.~J.,  {Bate} M.~R.,  2007, \mn@doi [\mnras]
  {10.1111/j.1365-2966.2007.11621.x}, \href
  {https://ui.adsabs.harvard.edu/abs/2007MNRAS.377...77P} {377, 77}

\bibitem[\protect\citeauthoryear{{Riaz}, {Bovino}, {Vanaverbeke}  \&
  {Schleicher}}{{Riaz} et~al.}{2018}]{2018MNRAS.479..667R}
{Riaz} R.,  {Bovino} S.,  {Vanaverbeke} S.,   {Schleicher} D.~R.~G.,  2018,
  \mn@doi [\mnras] {10.1093/mnras/sty1635}, \href
  {https://ui.adsabs.harvard.edu/abs/2018MNRAS.479..667R} {479, 667}

\bibitem[\protect\citeauthoryear{{Ripamonti} \& {Abel}}{{Ripamonti} \&
  {Abel}}{2004}]{2004MNRAS.348.1019R}
{Ripamonti} E.,  {Abel} T.,  2004, \mn@doi [\mnras]
  {10.1111/j.1365-2966.2004.07422.x}, \href
  {https://ui.adsabs.harvard.edu/abs/2004MNRAS.348.1019R} {348, 1019}

\bibitem[\protect\citeauthoryear{{Schekochihin}, {Cowley}, {Hammett}, {Maron}
  \& {McWilliams}}{{Schekochihin} et~al.}{2002}]{2002NJPh....4...84S}
{Schekochihin} A.~A.,  {Cowley} S.~C.,  {Hammett} G.~W.,  {Maron} J.~L.,
  {McWilliams} J.~C.,  2002, \mn@doi [New Journal of Physics]
  {10.1088/1367-2630/4/1/384}, \href
  {https://ui.adsabs.harvard.edu/abs/2002NJPh....4...84S} {4, 84}

\bibitem[\protect\citeauthoryear{{Schekochihin}, {Cowley}, {Taylor}, {Maron}
  \& {McWilliams}}{{Schekochihin} et~al.}{2004}]{2004ApJ...612..276S}
{Schekochihin} A.~A.,  {Cowley} S.~C.,  {Taylor} S.~F.,  {Maron} J.~L.,
  {McWilliams} J.~C.,  2004, \mn@doi [\apj] {10.1086/422547}, \href
  {https://ui.adsabs.harvard.edu/abs/2004ApJ...612..276S} {612, 276}

\bibitem[\protect\citeauthoryear{{Schleicher}, {Galli}, {Glover}, {Banerjee},
  {Palla}, {Schneider}  \& {Klessen}}{{Schleicher}
  et~al.}{2009}]{2009ApJ...703.1096S}
{Schleicher} D. R.~G.,  {Galli} D.,  {Glover} S. C.~O.,  {Banerjee} R.,
  {Palla} F.,  {Schneider} R.,   {Klessen} R.~S.,  2009, \mn@doi [\apj]
  {10.1088/0004-637X/703/1/1096}, \href
  {https://ui.adsabs.harvard.edu/abs/2009ApJ...703.1096S} {703, 1096}

\bibitem[\protect\citeauthoryear{{Schleicher}, {Banerjee}, {Sur}, {Arshakian},
  {Klessen}, {Beck}  \& {Spaans}}{{Schleicher}
  et~al.}{2010}]{2010A&A...522A.115S}
{Schleicher} D.~R.~G.,  {Banerjee} R.,  {Sur} S.,  {Arshakian} T.~G.,
  {Klessen} R.~S.,  {Beck} R.,   {Spaans} M.,  2010, \mn@doi [\aap]
  {10.1051/0004-6361/201015184}, \href
  {https://ui.adsabs.harvard.edu/abs/2010A&A...522A.115S} {522, A115}

\bibitem[\protect\citeauthoryear{{Schmidt}, {Federrath}, {Hupp}, {Kern}  \&
  {Niemeyer}}{{Schmidt} et~al.}{2009}]{2009A&A...494..127S}
{Schmidt} W.,  {Federrath} C.,  {Hupp} M.,  {Kern} S.,   {Niemeyer} J.~C.,
  2009, \mn@doi [\aap] {10.1051/0004-6361:200809967}, \href
  {https://ui.adsabs.harvard.edu/abs/2009A&A...494..127S} {494, 127}

\bibitem[\protect\citeauthoryear{{Schneider}, {Salvaterra}, {Ferrara}  \&
  {Ciardi}}{{Schneider} et~al.}{2006}]{2006MNRAS.369..825S}
{Schneider} R.,  {Salvaterra} R.,  {Ferrara} A.,   {Ciardi} B.,  2006, \mn@doi
  [\mnras] {10.1111/j.1365-2966.2006.10331.x}, \href
  {https://ui.adsabs.harvard.edu/abs/2006MNRAS.369..825S} {369, 825}

\bibitem[\protect\citeauthoryear{{Schober}, {Schleicher}, {Federrath},
  {Glover}, {Klessen}  \& {Banerjee}}{{Schober}
  et~al.}{2012}]{2012ApJ...754...99S}
{Schober} J.,  {Schleicher} D.,  {Federrath} C.,  {Glover} S.,  {Klessen}
  R.~S.,   {Banerjee} R.,  2012, \mn@doi [\apj] {10.1088/0004-637X/754/2/99},
  \href {https://ui.adsabs.harvard.edu/abs/2012ApJ...754...99S} {754, 99}

\bibitem[\protect\citeauthoryear{{Schober}, {Schleicher}, {Federrath}, {Bovino}
   \& {Klessen}}{{Schober} et~al.}{2015}]{2015PhRvE..92b3010S}
{Schober} J.,  {Schleicher} D.~R.~G.,  {Federrath} C.,  {Bovino} S.,
  {Klessen} R.~S.,  2015, \mn@doi [\pre] {10.1103/PhysRevE.92.023010}, \href
  {https://ui.adsabs.harvard.edu/abs/2015PhRvE..92b3010S} {92, 023010}

\bibitem[\protect\citeauthoryear{{Seiden}, {Schulman}  \& {Gerola}}{{Seiden}
  et~al.}{1979}]{1979ApJ...232..702S}
{Seiden} P.~E.,  {Schulman} L.~S.,   {Gerola} H.,  1979, \mn@doi [\apj]
  {10.1086/157329}, \href
  {https://ui.adsabs.harvard.edu/abs/1979ApJ...232..702S} {232, 702}

\bibitem[\protect\citeauthoryear{{Seifried}, {Banerjee}, {Klessen}, {Duffin}
  \& {Pudritz}}{{Seifried} et~al.}{2011}]{2011MNRAS.417.1054S}
{Seifried} D.,  {Banerjee} R.,  {Klessen} R.~S.,  {Duffin} D.,   {Pudritz}
  R.~E.,  2011, \mn@doi [\mnras] {10.1111/j.1365-2966.2011.19320.x}, \href
  {https://ui.adsabs.harvard.edu/abs/2011MNRAS.417.1054S} {417, 1054}

\bibitem[\protect\citeauthoryear{{Sharda}, {Krumholz}  \& {Federrath}}{{Sharda}
  et~al.}{2019}]{2019MNRAS.490..513S}
{Sharda} P.,  {Krumholz} M.~R.,   {Federrath} C.,  2019, \mn@doi [\mnras]
  {10.1093/mnras/stz2618}, \href
  {https://ui.adsabs.harvard.edu/abs/2019MNRAS.490..513S} {490, 513}

\bibitem[\protect\citeauthoryear{{Shu}, {Adams}  \& {Lizano}}{{Shu}
  et~al.}{1987}]{1987ARA&A..25...23S}
{Shu} F.~H.,  {Adams} F.~C.,   {Lizano} S.,  1987, \mn@doi [\araa]
  {10.1146/annurev.aa.25.090187.000323}, \href
  {https://ui.adsabs.harvard.edu/abs/1987ARA&A..25...23S} {25, 23}

\bibitem[\protect\citeauthoryear{{Sigl}, {Olinto}  \& {Jedamzik}}{{Sigl}
  et~al.}{1997}]{1997PhRvD..55.4582S}
{Sigl} G.,  {Olinto} A.~V.,   {Jedamzik} K.,  1997, \mn@doi [\prd]
  {10.1103/PhysRevD.55.4582}, \href
  {https://ui.adsabs.harvard.edu/abs/1997PhRvD..55.4582S} {55, 4582}

\bibitem[\protect\citeauthoryear{{Stacy} \& {Bromm}}{{Stacy} \&
  {Bromm}}{2014}]{2014ApJ...785...73S}
{Stacy} A.,  {Bromm} V.,  2014, \mn@doi [\apj] {10.1088/0004-637X/785/1/73},
  \href {https://ui.adsabs.harvard.edu/abs/2014ApJ...785...73S} {785, 73}

\bibitem[\protect\citeauthoryear{{Stacy}, {Greif}  \& {Bromm}}{{Stacy}
  et~al.}{2012}]{2012MNRAS.422..290S}
{Stacy} A.,  {Greif} T.~H.,   {Bromm} V.,  2012, \mn@doi [\mnras]
  {10.1111/j.1365-2966.2012.20605.x}, \href
  {https://ui.adsabs.harvard.edu/abs/2012MNRAS.422..290S} {422, 290}

\bibitem[\protect\citeauthoryear{{Stacy}, {Bromm}  \& {Lee}}{{Stacy}
  et~al.}{2016}]{2016MNRAS.462.1307S}
{Stacy} A.,  {Bromm} V.,   {Lee} A.~T.,  2016, \mn@doi [\mnras]
  {10.1093/mnras/stw1728}, \href
  {https://ui.adsabs.harvard.edu/abs/2016MNRAS.462.1307S} {462, 1307}

\bibitem[\protect\citeauthoryear{{Stevenson}}{{Stevenson}}{2003}]{2003E&PSL.208....1S}
{Stevenson} D.~J.,  2003, \mn@doi [Earth and Planetary Science Letters]
  {10.1016/S0012-821X(02)01126-3}, \href
  {https://ui.adsabs.harvard.edu/abs/2003E&PSL.208....1S} {208, 1}

\bibitem[\protect\citeauthoryear{{Subramanian}}{{Subramanian}}{1999}]{1999PhRvL..83.2957S}
{Subramanian} K.,  1999, \mn@doi [\prl] {10.1103/PhysRevLett.83.2957}, \href
  {https://ui.adsabs.harvard.edu/abs/1999PhRvL..83.2957S} {83, 2957}

\bibitem[\protect\citeauthoryear{{Subramanian}}{{Subramanian}}{2016}]{2016RPPh...79g6901S}
{Subramanian} K.,  2016, \mn@doi [Reports on Progress in Physics]
  {10.1088/0034-4885/79/7/076901}, \href
  {https://ui.adsabs.harvard.edu/abs/2016RPPh...79g6901S} {79, 076901}

\bibitem[\protect\citeauthoryear{{Sugimura}, {Matsumoto}, {Hosokawa}, {Hirano}
  \& {Omukai}}{{Sugimura} et~al.}{2020}]{2020ApJ...892L..14S}
{Sugimura} K.,  {Matsumoto} T.,  {Hosokawa} T.,  {Hirano} S.,   {Omukai} K.,
  2020, \mn@doi [\apjl] {10.3847/2041-8213/ab7d37}, \href
  {https://ui.adsabs.harvard.edu/abs/2020ApJ...892L..14S} {892, L14}

\bibitem[\protect\citeauthoryear{{Sur}, {Schleicher}, {Banerjee}, {Federrath}
  \& {Klessen}}{{Sur} et~al.}{2010}]{2010ApJ...721L.134S}
{Sur} S.,  {Schleicher} D.~R.~G.,  {Banerjee} R.,  {Federrath} C.,   {Klessen}
  R.~S.,  2010, \mn@doi [\apjl] {10.1088/2041-8205/721/2/L134}, \href
  {https://ui.adsabs.harvard.edu/abs/2010ApJ...721L.134S} {721, L134}

\bibitem[\protect\citeauthoryear{{Sur}, {Federrath}, {Schleicher}, {Banerjee}
  \& {Klessen}}{{Sur} et~al.}{2012}]{2012MNRAS.423.3148S}
{Sur} S.,  {Federrath} C.,  {Schleicher} D. R.~G.,  {Banerjee} R.,   {Klessen}
  R.~S.,  2012, \mn@doi [\mnras] {10.1111/j.1365-2966.2012.21100.x}, \href
  {https://ui.adsabs.harvard.edu/abs/2012MNRAS.423.3148S} {423, 3148}

\bibitem[\protect\citeauthoryear{{Sur}, {Pan}  \& {Scannapieco}}{{Sur}
  et~al.}{2014}]{2014ApJ...784...94S}
{Sur} S.,  {Pan} L.,   {Scannapieco} E.,  2014, \mn@doi [\apj]
  {10.1088/0004-637X/784/2/94}, \href
  {https://ui.adsabs.harvard.edu/abs/2014ApJ...784...94S} {784, 94}

\bibitem[\protect\citeauthoryear{{Susa}, {Hasegawa}  \& {Tominaga}}{{Susa}
  et~al.}{2014}]{2014ApJ...792...32S}
{Susa} H.,  {Hasegawa} K.,   {Tominaga} N.,  2014, \mn@doi [\apj]
  {10.1088/0004-637X/792/1/32}, \href
  {https://ui.adsabs.harvard.edu/abs/2014ApJ...792...32S} {792, 32}

\bibitem[\protect\citeauthoryear{{Susa}, {Doi}  \& {Omukai}}{{Susa}
  et~al.}{2015}]{2015ApJ...801...13S}
{Susa} H.,  {Doi} K.,   {Omukai} K.,  2015, \mn@doi [\apj]
  {10.1088/0004-637X/801/1/13}, \href
  {https://ui.adsabs.harvard.edu/abs/2015ApJ...801...13S} {801, 13}

\bibitem[\protect\citeauthoryear{{Tan}, {Beltr{\'a}n}, {Caselli}, {Fontani},
  {Fuente}, {Krumholz}, {McKee}  \& {Stolte}}{{Tan} et~al.}{2014}]{Tan14a}
{Tan} J.~C.,  {Beltr{\'a}n} M.~T.,  {Caselli} P.,  {Fontani} F.,  {Fuente} A.,
  {Krumholz} M.~R.,  {McKee} C.~F.,   {Stolte} A.,  2014, \mn@doi [Protostars
  and Planets VI] {10.2458/azu_uapress_9780816531240-ch007}, \href
  {http://adsabs.harvard.edu/abs/2014prpl.conf..149T} {pp 149--172}

\bibitem[\protect\citeauthoryear{{Tanaka}, {Tan}  \& {Zhang}}{{Tanaka}
  et~al.}{2017}]{2017ApJ...835...32T}
{Tanaka} K. E.~I.,  {Tan} J.~C.,   {Zhang} Y.,  2017, \mn@doi [\apj]
  {10.3847/1538-4357/835/1/32}, \href
  {https://ui.adsabs.harvard.edu/abs/2017ApJ...835...32T} {835, 32}

\bibitem[\protect\citeauthoryear{{Tashiro} \& {Sugiyama}}{{Tashiro} \&
  {Sugiyama}}{2006}]{2006MNRAS.372.1060T}
{Tashiro} H.,  {Sugiyama} N.,  2006, \mn@doi [\mnras]
  {10.1111/j.1365-2966.2006.10901.x}, \href
  {https://ui.adsabs.harvard.edu/abs/2006MNRAS.372.1060T} {372, 1060}

\bibitem[\protect\citeauthoryear{{Tseliakhovich} \& {Hirata}}{{Tseliakhovich}
  \& {Hirata}}{2010}]{2010PhRvD..82h3520T}
{Tseliakhovich} D.,  {Hirata} C.,  2010, \mn@doi [\prd]
  {10.1103/PhysRevD.82.083520}, \href
  {https://ui.adsabs.harvard.edu/abs/2010PhRvD..82h3520T} {82, 083520}

\bibitem[\protect\citeauthoryear{{Turk}, {Smith}, {Oishi}, {Skory}, {Skillman},
  {Abel}  \& {Norman}}{{Turk} et~al.}{2011}]{2011ApJS..192....9T}
{Turk} M.~J.,  {Smith} B.~D.,  {Oishi} J.~S.,  {Skory} S.,  {Skillman} S.~W.,
  {Abel} T.,   {Norman} M.~L.,  2011, \mn@doi [\apjs]
  {10.1088/0067-0049/192/1/9}, \href
  {http://adsabs.harvard.edu/abs/2011ApJS..192....9T} {192, 9}

\bibitem[\protect\citeauthoryear{{Turk}, {Oishi}, {Abel}  \& {Bryan}}{{Turk}
  et~al.}{2012}]{2012ApJ...745..154T}
{Turk} M.~J.,  {Oishi} J.~S.,  {Abel} T.,   {Bryan} G.~L.,  2012, \mn@doi
  [\apj] {10.1088/0004-637X/745/2/154}, \href
  {https://ui.adsabs.harvard.edu/abs/2012ApJ...745..154T} {745, 154}

\bibitem[\protect\citeauthoryear{{Turner} \& {Widrow}}{{Turner} \&
  {Widrow}}{1988}]{1988PhRvD..37.2743T}
{Turner} M.~S.,  {Widrow} L.~M.,  1988, \mn@doi [\prd]
  {10.1103/PhysRevD.37.2743}, \href
  {https://ui.adsabs.harvard.edu/abs/1988PhRvD..37.2743T} {37, 2743}

\bibitem[\protect\citeauthoryear{{Vachaspati}}{{Vachaspati}}{1991}]{1991PhLB..265..258V}
{Vachaspati} T.,  1991, \mn@doi [Physics Letters B]
  {10.1016/0370-2693(91)90051-Q}, \href
  {https://ui.adsabs.harvard.edu/abs/1991PhLB..265..258V} {265, 258}

\bibitem[\protect\citeauthoryear{{Vazza}, {Brunetti}, {Kritsuk}, {Wagner},
  {Gheller}  \& {Norman}}{{Vazza} et~al.}{2009}]{2009A&A...504...33V}
{Vazza} F.,  {Brunetti} G.,  {Kritsuk} A.,  {Wagner} R.,  {Gheller} C.,
  {Norman} M.,  2009, \mn@doi [\aap] {10.1051/0004-6361/200912535}, \href
  {https://ui.adsabs.harvard.edu/abs/2009A&A...504...33V} {504, 33}

\bibitem[\protect\citeauthoryear{{Vazza}, {Br{\"u}ggen}, {Gheller}  \&
  {Wang}}{{Vazza} et~al.}{2014}]{2014MNRAS.445.3706V}
{Vazza} F.,  {Br{\"u}ggen} M.,  {Gheller} C.,   {Wang} P.,  2014, \mn@doi
  [\mnras] {10.1093/mnras/stu1896}, \href
  {https://ui.adsabs.harvard.edu/abs/2014MNRAS.445.3706V} {445, 3706}

\bibitem[\protect\citeauthoryear{{Virtanen} et~al.,}{{Virtanen}
  et~al.}{2020}]{2020SciPy-NMeth}
{Virtanen} P.,  et~al., 2020, \mn@doi [Nature Methods]
  {https://doi.org/10.1038/s41592-019-0686-2}, \href {https://rdcu.be/b08Wh} {}

\bibitem[\protect\citeauthoryear{{Waagan}}{{Waagan}}{2009}]{2009JCoPh.228.8609W}
{Waagan} K.,  2009, \mn@doi [Journal of Computational Physics]
  {10.1016/j.jcp.2009.08.020}, \href
  {https://ui.adsabs.harvard.edu/abs/2009JCoPh.228.8609W} {228, 8609}

\bibitem[\protect\citeauthoryear{{Waagan}, {Federrath}  \&
  {Klingenberg}}{{Waagan} et~al.}{2011}]{2011JCoPh.230.3331W}
{Waagan} K.,  {Federrath} C.,   {Klingenberg} C.,  2011, \mn@doi [Journal of
  Computational Physics] {10.1016/j.jcp.2011.01.026}, \href
  {http://adsabs.harvard.edu/abs/2011JCoPh.230.3331W} {230, 3331}

\bibitem[\protect\citeauthoryear{{Wagstaff}, {Banerjee}, {Schleicher}  \&
  {Sigl}}{{Wagstaff} et~al.}{2014}]{2014PhRvD..89j3001W}
{Wagstaff} J.~M.,  {Banerjee} R.,  {Schleicher} D.,   {Sigl} G.,  2014, \mn@doi
  [\prd] {10.1103/PhysRevD.89.103001}, \href
  {https://ui.adsabs.harvard.edu/abs/2014PhRvD..89j3001W} {89, 103001}

\bibitem[\protect\citeauthoryear{{Widrow}, {Ryu}, {Schleicher}, {Subramanian},
  {Tsagas}  \& {Treumann}}{{Widrow} et~al.}{2012}]{2012SSRv..166...37W}
{Widrow} L.~M.,  {Ryu} D.,  {Schleicher} D. R.~G.,  {Subramanian} K.,  {Tsagas}
  C.~G.,   {Treumann} R.~A.,  2012, \mn@doi [\ssr] {10.1007/s11214-011-9833-5},
  \href {https://ui.adsabs.harvard.edu/abs/2012SSRv..166...37W} {166, 37}

\bibitem[\protect\citeauthoryear{{Wise} \& {Abel}}{{Wise} \&
  {Abel}}{2007}]{2007ApJ...665..899W}
{Wise} J.~H.,  {Abel} T.,  2007, \mn@doi [\apj] {10.1086/520036}, \href
  {https://ui.adsabs.harvard.edu/abs/2007ApJ...665..899W} {665, 899}

\bibitem[\protect\citeauthoryear{{Wise}, {Turk}  \& {Abel}}{{Wise}
  et~al.}{2008}]{2008ApJ...682..745W}
{Wise} J.~H.,  {Turk} M.~J.,   {Abel} T.,  2008, \mn@doi [\apj]
  {10.1086/588209}, \href
  {https://ui.adsabs.harvard.edu/abs/2008ApJ...682..745W} {682, 745}

\bibitem[\protect\citeauthoryear{{Wollenberg}, {Glover}, {Clark}  \&
  {Klessen}}{{Wollenberg} et~al.}{2020}]{2020MNRAS.494.1871W}
{Wollenberg} K. M.~J.,  {Glover} S. C.~O.,  {Clark} P.~C.,   {Klessen} R.~S.,
  2020, \mn@doi [\mnras] {10.1093/mnras/staa289}, \href
  {https://ui.adsabs.harvard.edu/abs/2020MNRAS.494.1871W} {494, 1871}

\bibitem[\protect\citeauthoryear{{W{\"u}nsch}, {Walch}, {Dinnbier}  \&
  {Whitworth}}{{W{\"u}nsch} et~al.}{2018}]{2018MNRAS.475.3393W}
{W{\"u}nsch} R.,  {Walch} S.,  {Dinnbier} F.,   {Whitworth} A.,  2018, \mn@doi
  [\mnras] {10.1093/mnras/sty015}, \href
  {https://ui.adsabs.harvard.edu/abs/2018MNRAS.475.3393W} {475, 3393}

\bibitem[\protect\citeauthoryear{{Wurster} \& {Li}}{{Wurster} \&
  {Li}}{2018}]{2018FrASS...5...39W}
{Wurster} J.,  {Li} Z.-Y.,  2018, \mn@doi [Frontiers in Astronomy and Space
  Sciences] {10.3389/fspas.2018.00039}, \href
  {https://ui.adsabs.harvard.edu/abs/2018FrASS...5...39W} {5, 39}

\bibitem[\protect\citeauthoryear{{Wurster}, {Bate}  \& {Price}}{{Wurster}
  et~al.}{2018}]{2018MNRAS.475.1859W}
{Wurster} J.,  {Bate} M.~R.,   {Price} D.~J.,  2018, \mn@doi [\mnras]
  {10.1093/mnras/stx3339}, \href
  {https://ui.adsabs.harvard.edu/abs/2018MNRAS.475.1859W} {475, 1859}

\bibitem[\protect\citeauthoryear{{Wurster}, {Bate}  \& {Price}}{{Wurster}
  et~al.}{2019}]{2019MNRAS.489.1719W}
{Wurster} J.,  {Bate} M.~R.,   {Price} D.~J.,  2019, \mn@doi [\mnras]
  {10.1093/mnras/stz2215}, \href
  {https://ui.adsabs.harvard.edu/abs/2019MNRAS.489.1719W} {489, 1719}

\bibitem[\protect\citeauthoryear{{Xu}, {O'Shea}, {Collins}, {Norman}, {Li}  \&
  {Li}}{{Xu} et~al.}{2008}]{2008ApJ...688L..57X}
{Xu} H.,  {O'Shea} B.~W.,  {Collins} D.~C.,  {Norman} M.~L.,  {Li} H.,   {Li}
  S.,  2008, \mn@doi [\apjl] {10.1086/595617}, \href
  {https://ui.adsabs.harvard.edu/abs/2008ApJ...688L..57X} {688, L57}

\bibitem[\protect\citeauthoryear{{Young} \& {Clarke}}{{Young} \&
  {Clarke}}{2016}]{2016MNRAS.455.1438Y}
{Young} M.~D.,  {Clarke} C.~J.,  2016, \mn@doi [\mnras]
  {10.1093/mnras/stv2378}, \href
  {https://ui.adsabs.harvard.edu/abs/2016MNRAS.455.1438Y} {455, 1438}

\bibitem[\protect\citeauthoryear{{Zweibel}}{{Zweibel}}{2006}]{2006AN....327..505Z}
{Zweibel} E.~G.,  2006, \mn@doi [Astronomische Nachrichten]
  {10.1002/asna.200610573}, \href
  {https://ui.adsabs.harvard.edu/abs/2006AN....327..505Z} {327, 505}

\makeatother
\end{thebibliography}


\bsp	
\label{lastpage}
\end{document}